\documentclass[aip,graphicx]{revtex4-1}
\draft 

\usepackage{graphicx}
\usepackage{dcolumn}
\usepackage{bm}

\begin{document}

\title{Exchange bias-like behavior due to hidden local magnetic state in a Weyl semimetal}

\author{Qing-Qi Zeng}
\affiliation{Anhui Key Laboratory of Low-energy Quantum Materials and Devices, High Magnetic Field Laboratory, HFIPS, Chinese Academy of Sciences, Hefei, Anhui 230031, China}

\author{Xi-Tong Xu}
\affiliation{Anhui Key Laboratory of Low-energy Quantum Materials and Devices, High Magnetic Field Laboratory, HFIPS, Chinese Academy of Sciences, Hefei, Anhui 230031, China}
\affiliation{Science Island Branch of Graduate School, University of Science and Technology of China, Hefei, 230026, China}

\author{En-Ke Liu}

\affiliation{Beijing National Laboratory for Condensed Matter Physics, Institute of Physics, Chinese Academy of Sciences, Beijing, 100190, P. R. China}

\author{Zhe Qu}

\affiliation{Anhui Key Laboratory of Low-energy Quantum Materials and Devices, High Magnetic Field Laboratory, HFIPS, Chinese Academy of Sciences, Hefei, Anhui 230031, China}
\affiliation{Science Island Branch of Graduate School, University of Science and Technology of China, Hefei, 230026, China}

\begin{abstract}
Magnetic Weyl semimetals, which couple magnetic order with topological features, have emerged as promising candidates for advanced topological-materials-based applications. The switching of magnetization and the driving of domain wall motion play key roles in developing such applications. In this study, we suggest that a type of hard-magnetic nuclei dominates the magnetic reversal and induces an exchange bias-like behavior with a prior magnetic history in the bulk Co$_3$Sn$_2$S$_2$ Weyl semimetal. The sign change of the exchange bias-like behavior can be realized by controlling the orientation of such hard-magnetic nuclei. Remarkably, these nuclei can retain their magnetic orientation at a temperature well above the Curie temperature of this material, suggesting the existence of a local magnetic state with non-zero magnetization and high stability. This local state is potentially related to the spin-orbit polaron reported in this system. Our study offers a new scenario for manipulating the magnetic reversal and offers new insights into the magnetism in this Weyl system.
\end{abstract}

\maketitle 

\section{Introduction\protect\\}
Time-reversal-symmetry broken Weyl semimetals have been widely studied in recent years\cite{RN830, RN838, RN837, RN1154, RN243, RN1178, RN1447, RN842, RN587, RN839, RN841, RN591}. One prominent example is Co$_3$Sn$_2$S$_2$, a hard-magnetic Weyl semimetal that crystallizes in a rhombohedral structure and features a Co–Sn Kagome layer\cite{RN504, RN243, RN505, RN587}.  It has a Curie temperature of approximately 175 K, a magnetic moment of $0.3\,\mu_B$ per Co, and an out-of-plane ($\mathit{c}$-axis) ferromagnetic ground state\cite{RN536, RN535, RN537} [Figs.~\ref{FIG_A}(a–b)]. Both large intrinsic anomalous Hall conductivity and anomalous Hall angle have been reported in single crystalline Co$_3$Sn$_2$S$_2$ bulk samples\cite{RN243, RN587}. Because of its topological properties and hard magnetism, Co$_3$Sn$_2$S$_2$ is widely studied in various fields such as topological thermal effects\cite{RN1225, RN1490, RN1500, RN1498},  spintronics\cite{RN1548, RN1443}, spin dynamics\cite{RN1541}, magneto-optical effect\cite{RN1229}, topological catalysis\cite{RN598}, and energy storage\cite{RN1538}.
\par
The topological properties of Co$_3$Sn$_2$S$_2$ are essentially related to its magnetism\cite{RN1411, RN1547, RN772, RN1537}. Hence, the magnetic properties of Co$_3$Sn$_2$S$_2$ including the ground magnetic state\cite{RN1540, RN535, RN1544}, magnetic domain behavior\cite{RN1418, RN1502, RN594}, and magnetic structure variations under different external conditions\cite{RN552, RN764, RN1546, RN1417} have attracted many interests. Because the manipulation of magnetization is central to spintronic applications, the magnetic reversal in this material is also widely studied. It has been reported that a D.C. current can modulate the coercivity and magnetic reversal in Co$_3$Sn$_2$S$_2$ nanoflakes at low current densities\cite{RN1443}. Besides, the exchange bias (EB) behavior, characterized by asymmetric positive and negative magnetic reversals has been reported in the Co$_3$Sn$_2$S$_2$ bulk samples\cite{RN884, RN1486}. Despite the unambiguously observed hysteresis loop shift (bias-like), whether it stems from the spin glasses or frustrated magnetic moments remains unclear. Generally, applying a positive initial magnetic field after zero-field-cooling (ZFC) will induce a negatively shifted loop. Notably, opposite shifts with the same initial direction after ZFC are observed in two measurements\cite{RN884}, which indicates an unusual manner of the EB-like behavior in this material.
\par
Hence, this study aims to clarify the factors driving the magnetic reversal of Co$_3$Sn$_2$S$_2$ bulk samples by investigating magnetic hysteresis loops with different prior set magnetic and thermal histories. It is found that a type of hard-magnetic nuclei dominates the magnetic reversal, whose influences require a few Teslas to eliminate despite the coercive field $\mathit{H_c}$ being the magnitude of 0.1 T. Consequently, not high enough maximum external magnetic field $\mathit{H_{max}}$ will leave some hard-magnetic nuclei, which can asymmetrically facilitate the magnetic reversals. The $\mathit{H_{max}}$-dependent $\mathit{H_c}$ offers another possible interpretation for the EB-like behavior in this material. More intriguingly, the hard-magnetic nuclei can even retain their orientations at temperatures much higher than the Curie temperature of 175 K in this system, indicating a robust local magnetic state.

\section{Method\protect\\}
Single-crystalline bulk samples were synthesized by a slow cooling method described in the literature\cite{RN824}. The orientation and composition were identified by X-ray diffraction and energy dispersive X-ray spectroscopy, respectively (see Supplemental Material Note. 1) for additional information on the as-grown sample). Magnetization measurements were conducted using a Quantum Design MPMS3-7 T system, and electric transport measurements were performed with a Janis 9 T magnet combined with a SynkTek MCL1-540 lock-in system. All measurements were conducted with the magnetic field aligned along the $\mathit{c}$-axis, perpendicular to the Kagome lattice.

\section{Results and Discussion\protect\\}
\subsection{Tunable EB-like behavior, $\mathit{H_{max}}$-dependent $\mathit{H_c}$}
As aforementioned, the spontaneous EB-like behavior (after ZFC) was inconsistent across measurements, prompting us to adopt a field-cooling (FC) approach. The samples were cooled to 5 K under a magnetic field of $-0.02$ T. Next, the external magnetic field was initially applied to $+0.05$ T, and then a Hall hysteresis loop with $\mathit{H_{max}^\pm}$ = $\pm 0.05$ T was measured. Hysteresis loops were measured with gradually increasing $\mathit{H_{max}^\pm}$ (up to $\pm5$ T).
 \par
\begin{figure*}[htbp]
  \begin{center}
  \includegraphics[clip, width=17 cm]{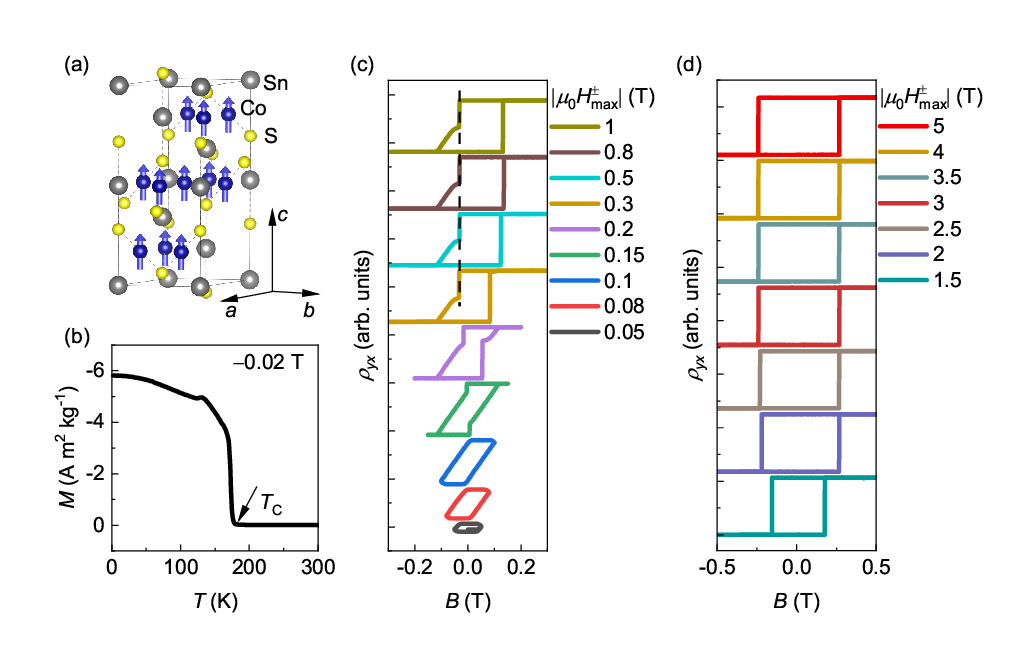}\\[1pt]
  \caption{
  (a) Crystal structure of Co$_3$Sn$_2$S$_2$. Arrows illustrate magnetic moments of Co atoms.
  (b) Temperature-dependent magnetization measured under a magnetic field of $-0.02$ T.
  (c-d) Hall loops with gradually increased maximum external magnetic field $\mathit{H_{max}^\pm}$ at 5 K for sample \#a. Dashed black line provides visual guidance for the unchanged negative coercive field $\mathit{H_c^-}$.
  }
  \label{FIG_A}
  \end{center}
\end{figure*}

 As Figs.~\ref{FIG_A}(c–d) shows, when $\mathit{|H_{max}^\pm|}$ is less than 0.1 T, the loops are linear-increased and hysteretic, and anti-symmetric about the zero field. That is, no shift of the loop is observed. Observable loop shifts appear as $\mathit{|H_{max}^\pm|}$ ranges from 0.3 to 1 T, which is reminiscent of EB. In these loops, the positive reversals are in a single jump and the negative ones follow a two-step way, i.e., a sudden drop followed by a linear segment at the same critical field. As for $\mathit{|H_{max}^\pm|}$ higher than 1 T, both the positive and negative reversals occur abruptly, and the loop shift disappears at $\mathit{|H_{max}^\pm|}=3$ T. Despite the different critical fields, sample \#b shows similar behavior as sample \#a. The magnetization loops of sample \#b with the same protocol also show consistent results with the Hall curves (Supplemental Material Note. 2).
 \par

\begin{figure}[htbp]
  \begin{center}
  \includegraphics[clip, width=6 cm]{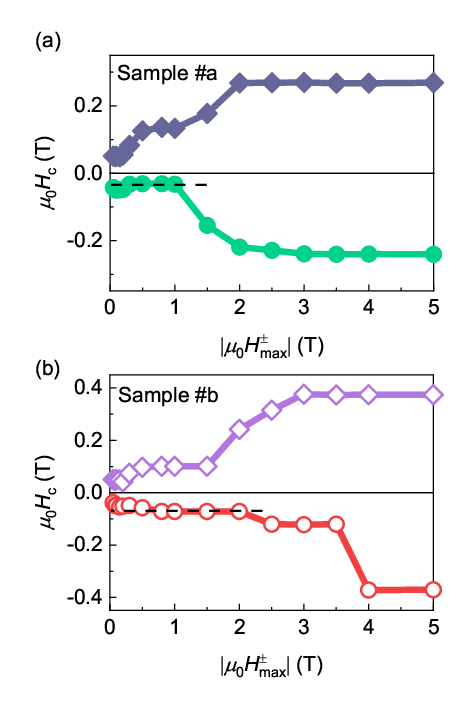}\\[1pt]
    
  \caption{
    $\mathit{H_{max}}$-dependent $\mathit{H_c^\pm}$ for samples \#a and \#b. Dashed black lines provide visual guidance for the almost unchanged $\mathit{H_c^-}$.
    }
  \label{FIG_B}
  \end{center}
\end{figure}

The above results show that $\mathit{H_c}$ in Co$_3$Sn$_2$S$_2$ can be affected by $\mathit{H_{max}}$. Asymmetric positive and negative reversals, that is, an EB-like behavior appearing at a moderate $\mathit{H_{max}}$ but vanishing under high enough $\mathit{H_{max}}$. In addition, as Fig.~\ref{FIG_B} shows, $\mathit{H_c^\pm}$ increase asymmetrically in a step-like way when $\mathit{H_{max}}$ is over some critical values. For example, $\mathit{H_c^-}$ keeps a value lower than $-0.1$ T when $\mathit{|H_{max}^\pm|}$ is between 0.3 and 1 T for sample \#a, and the maximum value of $\mathit{H_c^-}$ requires a $\mathit{|H_{max^\pm|}}$ of 4 T for sample \#b.

\begin{figure*}[htbp]
  \begin{center}
  \includegraphics[clip, width=17 cm]{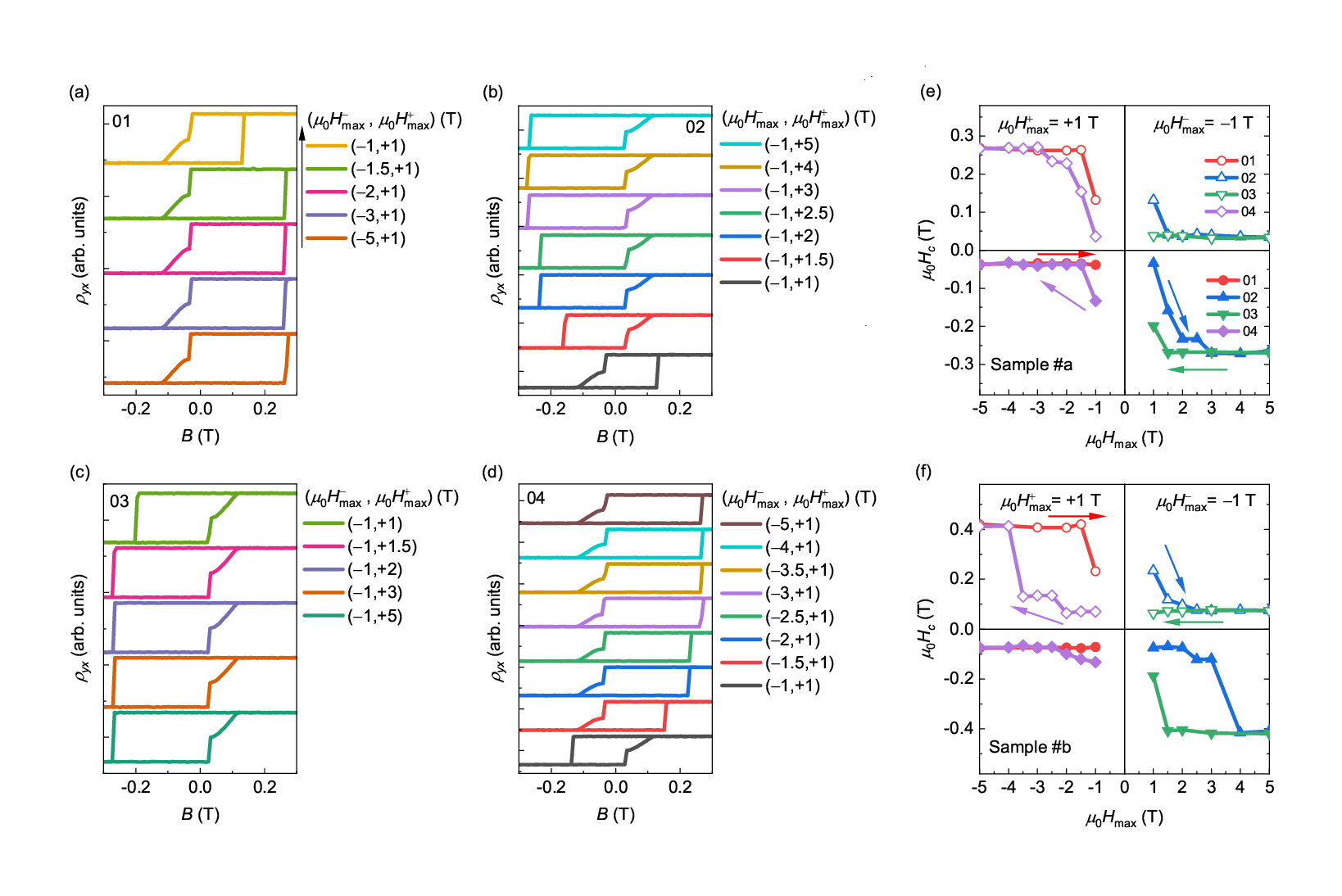}\\[1pt]
  
  \caption{
      (a–d) Asymmetric Hall loops with different $\mathit{H_{max}^+}$ and $\mathit{H_{max}^-}$ (listed in legend) at 5 K for sample \#a. Protocols 01–04 were performed continuously. (e–f) Coercive fields in protocols 01–04. In each figure, left panel shows the results with $\mathit{H_{max}^+}$ $= +1$ T and varied $\mathit{H_{max}^-}$, while right panel shows an opposite protocol. Up and down panels in each figure show positive and negative coercive fields, respectively.}
  \label{FIG_C}
  \end{center}
\end{figure*}

With the realization of the close relation between $\mathit{H_c}$ and $\mathit{H_{max}}$, the tunability of the magnetic reversal was further studied by performing a protocol with asymmetric $\mathit{H_{max}^\pm}$. Results of sample \#a are shown following and those of sample \#b can be found in the Supplemental Material Note. 3. After the final sweep in Fig.~\ref{FIG_A}(d), the external magnetic field was oscillated and decreased to zero, and the sample was demagnetized in this way. And then, we started with a negative saturated but positive non-saturated state. Specifically, series Loops with asymmetric $\mathit{H_{max}^\pm}$ were measured for both samples. In protocol 01, $\mathit{H_{max}^-}$ decreased from $-5$ to $-1$ T while $\mathit{H_{max}^+}$ kept $+1$ T in each sweep. Next, $\mathit{H_{max}^-}$ kept $-1$ T while $\mathit{H_{max}^+}$ increased from $+1$ to $+5$ T in protocol 02. The magnetic history in protocols 03 and 04 was analogous but anti-symmetric with that in protocols 01 and 02. All the measurements were continuously performed, as shown in Fig.~\ref{FIG_C}, where the $\mathit{H_{max}^\pm}$ for each loop were listed in the legend.

Recall that the final loop in Fig.~\ref{FIG_A}(d) with $\mathit{H_{max}^\pm}$ $= \pm5$ T has no shift. Nevertheless, as Fig.~\ref{FIG_C}(a) shows, EB-like behavior appears again by applying asymmetric $\mathit{H_{max}^\pm}$ without the requirement of rewarming and FC. For all loops in protocol 01, the positive reversals appear in a single jump and $\mathit{H_c^+}$ retains its maximum value till $\mathit{H_c^-}$ decrease to $-1.5$ T. When $\mathit{H_c^-}$ decreases to $-1$ T, $\mathit{H_c^+}$ is lower than the maximum value. In contrast, every negative reversal shows a sudden drop at the same critical field followed by a linear segment. Owing to this reversal at a very low field, all the loops in protocol 01 are positively shifted.
\par
In protocol 02 [Fig.~\ref{FIG_C}(b)], the reversal at a very low magnetic field appears in the positive direction when $\mathit{H_c^+}$ increases to $+1.5$ T. After that, the loops are negatively shifted, and $\mathit{H_c^-}$ increases to its maximum value gradually with increasing $\mathit{H_{max}^+}$. Besides, the first loop in protocol 01 and the last loop in protocol 02 are anti-symmetric, which shows the EB-like behavior is completely reversed by varying $\mathit{H_{max}^\pm}$. Owing to the similar but anti-symmetric magnetic histories, the Hall hysteresis loops in protocols 03 and 04 vary in an anti-symmetric manner with that in protocols 01 and 02. Sample \#b shows analogous behavior with different critical fields (Supplemental Material Note. 3).
\par
The variation of $\mathit{H_c^\pm}$ in protocols 01–04 is shown in Figs.~\ref{FIG_C}(e–f). One can find that both the positive and negative coercive fields vary in a hysteretic way. This indicates that the responses of the key factors (resulting in the observed shift) to the asymmetric $\mathit{H_{max}^\pm}$ conditions have a hysteresis. In addition, the tunability of the EB-like behavior can be realized by specific prior magnetic history.
\par
Thermal activation is also an external condition that can affect the magnetic behavior including the magnetic order, domain, and reversal in a material. Hence, we further identified the temperature dependence of the asymmetric magnetic reversal in this system. Hall loops with a positively saturated but negatively unsaturated protocol ($\mathit{H_{max}^+}$ $= +5$ T and $\mathit{H_{max}^-}$ $= -1$ T) were measured at different temperatures. Some of the loops are shown in Fig.~\ref{FIG_D}(a), more details can be found in Supplemental Material Note. 4. As $\mathit{H_{max}^+}$ of $+5$ T is sufficiently large, $\mathit{H_c^-}$ should be its maximum value at all temperatures. As Fig.~\ref{FIG_D}(a) and Fig.~\ref{FIG_D}(b) show, $\mathit{H_c^-}$ decreases gradually with increasing temperature while $\mathit{H_c^+}$ show a minor temperature dependence below 70 K (60 K) for sample \#a (sample \#b).
\par
When the temperature is increased to 75 K (65 K for sample \#b), $\mathit{H_c^+}$ increases and the difference between the $\mathit{H_c^\pm}$ ($\Delta\mathit{H_c} = \mathit{|H_c^+|} - \mathit{|H_c^-|}$) decreases suddenly [Figs. \ref{FIG_D}(b–c)], whereas $\mathit{H_c^+}$ is still lower than $\mathit{H_c^-}$. In both samples, $\Delta\mathit{H_c}$ decreases to zero at 95 K. The kind of curve with a sudden drop followed by a linear segment shows again above 110 K, but in an anti-symmetric manner in both the positive and negative reversals.

\begin{figure*}[htbp]
  \begin{center}
  \includegraphics[clip, width=17 cm]{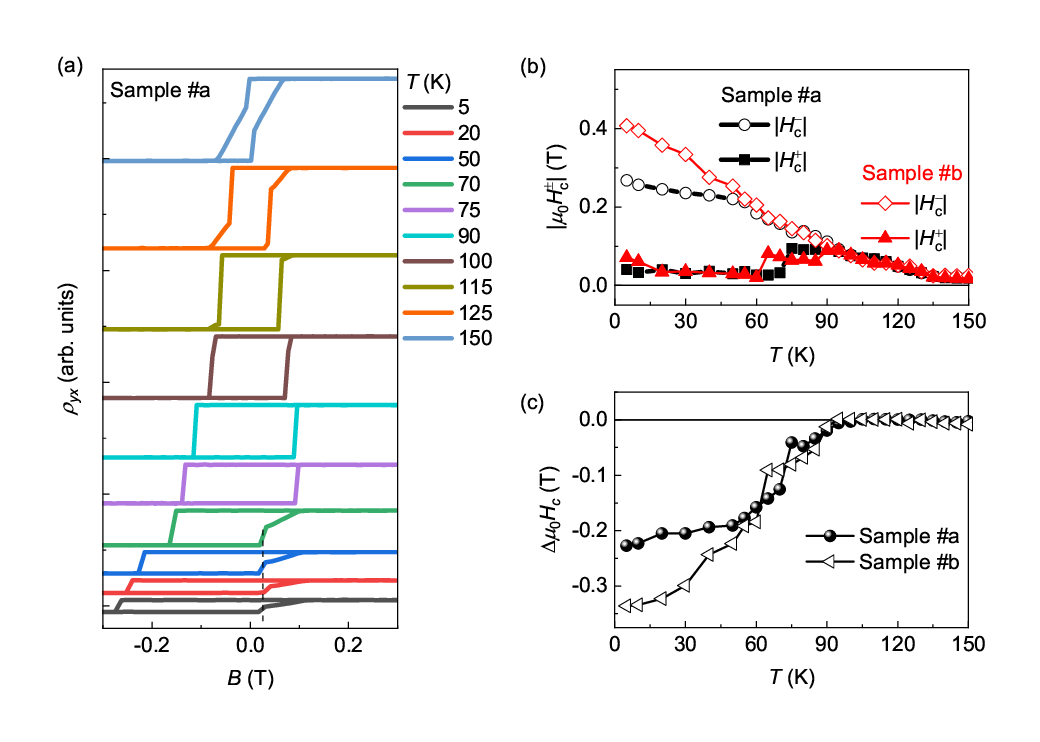}\\[1pt]
  
  \caption{
      (a) Asymmetric Hall loops with $\mathit{H_{max}^+}$ $= +5$ T and $\mathit{H_{max}^-}$ $= -1$ T at different temperatures for sample \#a. (b) Temperature-dependent $\mathit{|H_c^\pm|}$. (c) Difference between $\mathit{H_c^\pm}$ ($\Delta\mathit{H_c} = \mathit{|H_c^+|} - \mathit{|H_c^-|}$) versus temperature.}
  \label{FIG_D}
  \end{center}
\end{figure*}

\par
According to the above results, the magnetic reversal in the Co$_3$Sn$_2$S$_2$ system can occur asymmetrically with a moderate $\mathit{H_{max}}$ or asymmetrically applied $\mathit{H_{max}^\pm}$ at low temperatures. The shift observed in the hysteresis loop is analogous to an EB effect. In addition, the shift direction can be controlled by varying $\mathit{H_{max}}$ without a rewarming combined FC process. Moreover, high enough $\mathit{H_{max}}$ will eliminate the shift of the loop, and thermal activation also smears the EB-like behavior. The abovementioned tunability and the elimination of the EB-like behavior can be realized by just varying $\mathit{H_{max}}$, which indicates a difference between the observed phenomenon and the conventional EB effect.

\subsection{Nuclei dominated magnetic reversal in Co$_3$Sn$_2$S$_2$} 

The observed EB-like behavior can be explained based on the coercivity mechanisms. Magnetic hysteresis loops in real materials are affected by nucleation, domain-wall motion, and rotation of the moments. The last one can be a dominant factor in a very small particle with a single domain. In contrast, for a bulk, its magnetic reversal is usually dominated by a multidomain mechanism involving the formation of reverse domains and the DW pinning\cite{RN1520, RN1521, RN1527, RN1530}. In a magnet whose reversal is dominated by the DW pinning, the DWs are pinned by a dispersion of many inhomogeneities, especially planar defects such as grain boundary\cite{RN1520, RN1518}, which is expected to have a minor influence in a single crystal.
\par
For single crystals, the influence of nucleation is more pronounced. A reverse domain can form in two ways. One is the depinning of an already existing small region with an opposite magnetization to the majority, i.e., a sub-domain that is pinned by one or very few defects (no need for planner defects)\cite{RN1518}. The sub-domains may originate from the fragments left by DW moving at local pinning sites. These sub-domains can survive after a magnetic reversal if the external magnetic field is not applied large enough. Then the next reversal can be facilitated by the left sub-domains, whose depinning field may be a relatively low value. That is why surface smoothness is employed to reduce pinning sites and enhance the coercivity in permanent magnets\cite{RN1530, RN1524}. These sub-domains are named type \uppercase\expandafter{\romannumeral1} nuclei in the previous literature\cite{RN1518, RN1529, RN1524}. The other way to form reverse domains involves the real nucleation process, in which the reverse domains form at critical fields in a completely saturated sample. For example, large local demagnetizing fields in corners or pits, and inhomogeneities with different magnetic anisotropy may aid the formation of small reverse domains at some specific places\cite{RN1520, RN1518, RN1551, RN1530, RN1523, RN1550}. These type \uppercase\expandafter{\romannumeral2} nuclei are always present at the same places with a certain magnitude of critical fields for either direction of the magnetic reversal. For both types of nuclei, the nucleus with the smallest critical field will expand the most easily and dominate the magnetic reversal. For convenience, all the ‘nuclei’ without specific mention refer to type \uppercase\expandafter{\romannumeral1} nuclei in this text.
\par
In the studied system [see Figs.~\ref{FIG_A}(b–c) and Fig. S2], when $\mathit{|H_{max}^\pm|}$ is less than 0.1 T, the loops are linear and hysteretic. The linear segment in the magnetization is usually attributed to a geometric demagnetization effect\cite{RN1518, RN1523}. Nevertheless, it is not the only reason for the Co$_3$Sn$_2$S$_2$ system. The demagnetization factor is calculated\cite{RN1406} and the maximum demagnetization field is estimated to be 0.07 T for both samples (Supplemental Material Note. 5), which is less than 0.1 T. Moreover, the visible hysteresis related to the energy loss also shows that the curve does not simply result from a geometric demagnetization effect. The hysteresis loops should be affected by both the demagnetization effect and the pinning of DWs. The largest DW pinning field is approximately 0.1 T.
\par
When $\mathit{|H_{max}^\pm|}$ is higher than 0.1 T, sudden drops related to type \uppercase\expandafter{\romannumeral1} nuclei\cite{RN1520, RN1518, RN1523} are present in the reversals. Every critical field for a type \uppercase\expandafter{\romannumeral1} nucleus to reverse the magnetization is related to its depinning field and is marked as $\mathit{H_n}$. It should be kept in mind that in real systems, the nuclei and pinning effects always coexist. DWs will be pinned if the external magnetic field cannot offer enough energy to overcome the energy barrier. Hence, the reversal will be a sudden drop followed by a linear segment if $\mathit{H_n}$ is less than the largest DW pinning field in a sample. Once $\mathit{|H_n^\pm|}$ increases to higher than 0.1 T, the energy barrier can be overcome concurrently with the nuclei process therefore the DWs will sweep out the whole sample rapidly, and the magnetic reversal occurs in a single jump.
\par
The least trapped type \uppercase\expandafter{\romannumeral1} nucleus will expand first to reverse the magnetization. If the nucleus with the lowest $\mathit{H_n}$ is moved out by the external magnetic field, other nuclei with higher $\mathit{H_n}$ will lead to higher $\mathit{H_c}$ in the next reversal, which is the reason for the increased $\mathit{H_c}$ with increasing $\mathit{H_{max}}$. If all the type \uppercase\expandafter{\romannumeral1} nuclei are dislodged by the external field, the following magnetic reversals will be determined by the real nucleation process, i.e., the type \uppercase\expandafter{\romannumeral2} nuclei. As the type \uppercase\expandafter{\romannumeral2} nuclei are always present at the same places for either direction of the reversal, the maximum $\mathit{|H_c^+|}$ has the same value as $\mathit{|H_c^-|}$.
\par
The depinning and the dislodging fields for a nucleus can differ by an order of magnitude\cite{RN1518, RN1523, RN1524}. That is why the magnetization reversal starts below 0.5 T but the real saturation of this system needs a few Teslas. Similar behavior has been reported in permanent magnets\cite{RN1525, RN1524}. In addition, the nuclei originating from the fragments left by the DW moving may not have the same distributions for every reversal\cite{RN1523}. That can explain why $\mathit{H_c^\pm}$ are not anti-symmetric in those loops with moderate $\mathit{H_{max}}$.

\subsection{Controllable hard-magnetic nuclei}
Although the asymmetrically increased $\mathit{H_c}$ with increasing $\mathit{H_{max}}$, and the saturation of $\mathit{|H_c^\pm|}$ values under high enough $\mathit{H_{max}}$ in the Co$_3$Sn$_2$S$_2$ system can be understood based on the coercivity mechanisms, there is still a concern that the asymmetrically distributed fragments can induce a different $\mathit{|H_c^\pm|}$ but may not be responsible for the tunability of the EB-like behavior.
\par
It is noticed that the sudden drop at a field lower than 0.1 T [see Fig.~\ref{FIG_C} for example] plays a key role in the EB-like behavior under moderate $\mathit{H_{max}}$ in this system. The shift of the loop will reverse concurrently with the sign change of this reversal. In addition, the critical fields of this reversal are anti-symmetric in the positive and negative directions, and the critical $\mathit{H_{max}}$ realizing the sign change of this reversal is 1.5 T (2.5 T) for sample \#a (sample \#b) in both directions. All the abovementioned results indicate the existence of a type of nuclei that can be switched by an external field.
\par
We suggest that besides the type \uppercase\expandafter{\romannumeral1} nuclei resulting from the unreproducible fragments, there are hard-magnetic nuclei in the Co$_3$Sn$_2$S$_2$ system. They behave like discrete magnetic moments and can keep their orientations opposite with the magnetization if $\mathit{H_{max}}$ is not high enough, and therefore can aid the next magnetic reversal. Every hard-magnetic nucleus has its $\mathit{H_n}$ and flipping field. Although it is difficult to observe the characteristics of all the hard-magnetic nuclei according to the macro-measurements, we can still recognize two special ones: one that determines the very low $\mathit{H_n}$ is marked as $\mathit{n}$(0), and the other with the highest flipping field is marked as $\mathit{n}$(1). The depinning and flipping fields of $\mathit{n}$(0) and $\mathit{n}$(1) for the two samples are collected from Figs.~\ref{FIG_C}(e–f) and listed in Table~\ref{T1}.

\begin{table}[htbp]
  \footnotesize
  \caption{\label{T1}
  Depinning ($\mathit{H_n}$) and flipping fields ($\mathit{H_f }$) of $\mathit{n}$(0) and $\mathit{n}$(1).
  }
  \begin{tabular}{ccccc}
    \hline
      & \multicolumn{2}{c}{$\mathit{n}$(0)} & \multicolumn{2}{c}{$\mathit{n}$(1)} \\ \hline
      & ~$\mu_0\mathit{H_n}$ (T)~           & ~$\mu_0\mathit{H_f}$ (T)~          & ~~~~~$\mu_0\mathit{H_n}$ (T)~            & ~$\mu_0\mathit{H_f}$ (T)~         \\\hline
      Sample \#a~ & 0.03        & 1.5        & 0.23         & 3         \\
      Sample \#b~ & 0.08        & 2.5        & 0.12         & 4         \\ \hline
    \end{tabular}
\end{table}

Based on this assumption, the tunability of the EB-like behavior can be understood. In Fig.~\ref{FIG_B}(a), because the magnetic field of $-5$ T is sufficient to saturate the majority of the sample, all the sub-domains, and the hard-magnetic nuclei, a maximum $\mathit{H_c^+}$ determined by type \uppercase\expandafter{\romannumeral2} nuclei is observed until $\mathit{H_{max}^-}$ $= -1.5$ T. When $\mathit{H_{max}^-}$ decreases to $-1$ T, $\mathit{H_c^+}$ is lower than its maximum value. However, that does not imply a release of any hard-magnetic nuclei. The reason should be that the fragments are present in every reversal\cite{RN1518, RN1523}, not high enough $\mathit{H_c^-}$ will leave some of them that can be the type \uppercase\expandafter{\romannumeral1} nuclei and lower the $\mathit{H_c^+}$ in the next reversal. In addition, because $\mathit{H_{max}^-}$ of +1 T cannot flip $\mathit{n}$(0) to the positive direction, all the negative reversals are dominated by $\mathit{n}$(0)’s low $\mathit{H_n}$ in protocol 01.
\par
In protocol 02 [Fig.~\ref{FIG_B}(b)], when $\mathit{H_{max}^+}$ increases to +1.5 T, the direction of $\mathit{n}$(0)’s moment is flipped by the sufficient magnetic field and remains in the positive direction as $\mathit{H_{max}^-}$ keeps $-1$ T in the following. As a result, $\mathit{n}$(0) starts to aid the positive reversal, and the shift of the loop changes its sign. In addition, $\mathit{n}$(1), the hard-magnetic nucleus with the highest flipping field, remains in the negative direction until $\mathit{H_{max}^+}$ increases to $+3$ T. After that, $\mathit{n}$(1) is aligned to the positive direction, and therefore the negative reversal is determined by type \uppercase\expandafter{\romannumeral2} nuclei and $\mathit{H_c^-}$ reaches its maximum value. The first loop in protocol 01 and the last loop in protocol 02 are anti-symmetric owing to the $\mathit{n}$(0) and type \uppercase\expandafter{\romannumeral2} nuclei dominated negative and positive reversals in protocol 01, and the opposite cases in protocol 02.
\par
The magnetic hysteresis loop in protocols 03 and 04 can be understood similarly. Whereas, we notice that $\mathit{|H_c^-|}$ of the last loop in protocol 03 and $\mathit{|H_c^+|}$ of the last loop in protocol 01 are not equal. As the external magnetic field of $\pm1$ T is not high enough to eliminate all the fragments, the different $\mathit{H_c}$ mentioned above may be due to the unproducible distribution of the fragments.
\par
In contrast to the fragments, the hard-magnetic nuclei keep their orientations during a magnetic reversal till the external magnetic field is sufficient to flip them. In other words, the hard-magnetic nuclei such as $\mathit{n}$(0) and $\mathit{n}$(1) can be flipped but cannot be dislodged by an external magnetic field. Hence, they always show consistent depinning and flipping fields in protocols 01–04. That is the main difference between these two types of type \uppercase\expandafter{\romannumeral1} nuclei.

\subsection{Hard-magnetic nuclei versus thermal activation}

The above discussions show the existence and the tunability of the hard-magnetic nuclei that dominate the magnetic reversal with prior magnetic history in Co$_3$Sn$_2$S$_2$. We further interpret the temperature dependence of the magnetic reversal observed in Fig.~\ref{FIG_D}.
\par
As $\mathit{H_{max}^+}$ of $+5$ T is sufficient to saturate the sample, $\mathit{H_{max}^-}$ is dominated by type \uppercase\expandafter{\romannumeral2} nuclei at all temperatures. The critical field for type \uppercase\expandafter{\romannumeral2} nuclei is lower at higher temperatures. In addition, $\mathit{H_{max}^+}$ is determined by $\mathit{n}$(0) at low temperatures, which shows a minor temperature dependence below 70 K (60 K) for sample \#a (sample \#b). The suddenly increased $\mathit{H_{max}^+}$ at 75 K for sample \#a (65 K for sample \#b) [Fig. \ref{FIG_D}(b)] will be interpreted later.
\par
As the type \uppercase\expandafter{\romannumeral2} nuclei dominated $\mathit{H_{max}^-}$ monotonously decreases with increasing temperature, the equal absolute values of $\mathit{H_{max}^\pm}$ above 95 K are attributed to that the formation of the type \uppercase\expandafter{\romannumeral2} nuclei has the lowest critical field in this temperature range. For the same reason, the sudden drop above 110 K is also attributed to the type \uppercase\expandafter{\romannumeral2} nuclei, and the linear increased curve is related to the demagnetizing and pinning effects. Although the shape of the curve is analogous to that below 70 K, the origin of the sudden drop is different.
\par
In the following, we concentrate on the temperature-dependent behavior of the hard-magnetic nuclei. There are two possibilities for the unobservable hard-magnetic nuclei at high temperatures. One is that they are deactivated by thermal activation, i.e., disappear at high temperatures. The other possibility is that their flipping field decreases to lower than 1 T, in which case they are aligned to the direction of the magnetization after every reversal (in the studied protocol) and therefore do not aid the next reversal.
\par
\begin{figure*}[htbp]
  \begin{center}
  \includegraphics[clip, width=17 cm]{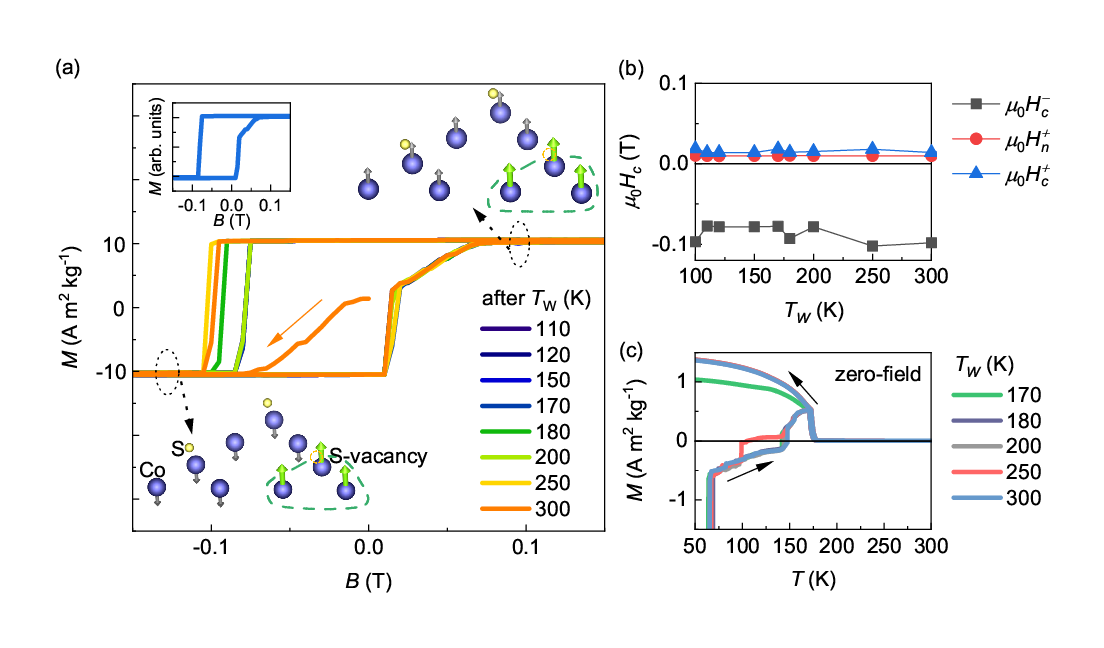}\\[1pt]
  
  \caption{
    Magnetization with prior set magnetic history of $+1$ T at 100 K for sample \#b. (a) Hysteresis loop with $\mathit{H_{max}^\pm\,=\,\pm}$1 T at 50 K after zero-field-cooling (ZFC) from 100 K. (b) Temperature dependence of magnetization in a zero-field-warming (ZFW) and ZFC process. (c) Hysteresis loop with $\mathit{H_{max}^\pm\,=\,\pm}$1 T at 50 K after ZFW and ZFC processes with warming temperature $\mathit{T_W\,=\,}$110 K. Loops at 50 K (d), extracted critical fields at 50 K (e), and temperature-dependent magnetization with different $\mathit{T_W}$ in cycling measurements. Every hysteresis loop in (a), (c), and (d) starts from a negative magnetic field of $-1$ T. Inset of (e) sketches the local magnetic state. Blue and yellow are Co and S atoms, and Sn atoms are not shown for clear display.}
  \label{FIG_E}
  \end{center}
\end{figure*}

Among those hard-magnetic nuclei, $\mathit{n}$(0) is easily to be observed. As aforementioned, the influence of $\mathit{n}$(0) disappears at 75 K (65 K for sample \#b). To figure out the influence of the thermal activation on $\mathit{n}$(0), a protocol that includes zero-field warming (ZFW), ZFC, and the measurements of isothermal magnetic hysteresis loops is performed. First, a magnetic field of $+1$ T was applied to sample \#b at 100 K (higher than the critical temperature mentioned above). Next, the sample was cooled to 50 K under zero field. After that, a hysteresis loop starting from $-1$ T with $\mathit{H_{max}^\pm\,=\,\pm}$1 T was measured [shown in Fig.~\ref{FIG_E}(a)]. A negatively shifted hysteresis loop is observed.

\par
In the following, the sample was warmed to different temperatures and cooled to 50 K under zero-field, after which isothermal loops at 50 K were measured. The temperature dependence of the magnetization and the isothermal hysteresis loop with a warming temperature $\mathit{T_W}$ of 110 K are shown in Figs.~\ref{FIG_E}(b–c) as example. Figure~\ref{FIG_E}(d) shows other loops with different $\mathit{T_W}$, where the initial magnetization curve for $\mathit{T_W\,=\,}$300 K is shown. 
\par
Despite the discordance negative reversals, all the positive reversals follow the same path, and all $\mathit{H_n^+}$ and $\mathit{H_c^+}$ are the same values [Figs.~\ref{FIG_E}(d–e)]. It should originate from the facilitation to the positive reversal from $\mathit{n}$(0). The initial magnetized direction at 50 K was negative. Additionally, the remanence was also negative while the external magnetic field was decreased to zero from $-1$ T after the measurement of every loop. Both the signs of the initial field and the remanence in the previous loop are negative, so the positively aligned $\mathit{n}$(0) is a memory of the magnetic history of $+1$ T at 100 K. As $\mathit{n}$(0) is not the nucleus with the highest flipping field at low temperatures, one can expect more hard-magnetic nuclei existing at 300 K in this system.
\par
The temperature-dependent magnetization measured in the ZFW process shows a sign change at a moderate temperature [Fig.~\ref{FIG_E}(f)], which should be due to the competition between the negative remanence and the positively orientated hard-magnetic nuclei. Furthermore, although the magnetization decreases to zero at the Curie temperature of 175 K, the sample is spontaneously magnetized to the positive direction in the following ZFC process. These results show the ability of the hard-magnetic nuclei to magnetize the sample.
\par
Intriguing information can be concluded from the above results. The reason for the unobservable influence of $\mathit{n}$(0) in Fig.~\ref{FIG_D}(a) is that $\mathit{n}$(0)’s flipping field decreases to lower than 1 T above 70 K (60 K) for sample \#a (sample \#b), not the deactivation of $\mathit{n}$(0) itself. In addition, the hard-magnetic nuclei can magnetize the sample and can retain their orientation even when the sample is zero-field warmed to 300 K. 
\par
Now one can understand why the hysteresis loops after ZFC show inconsistent positive and negative shifts in two measurements\cite{RN884}. There might be an unrecorded magnetic history in the authors’ measurements.
\par
We suggest that the hard-magnetic nuclei in Co$_3$Sn$_2$S$_2$ are local magnetic states with non-zero net magnetization and can align the moments near it by the exchange interaction. Strong local moments (sketched in the inset of Fig.~\ref{FIG_E}(d)) have been reported previously, which originate from the localized spin-orbit polarons nucleated around S-vacancies in Co$_3$Sn$_2$S$_2$\cite{RN1230}. As reported, the bulk S-vacancies can enhance the magnetic moment of neighboring Co atoms. This supports the existence of the discrete magnetic state in the Co$_3$Sn$_2$S$_2$ system.

\section{Summary\protect\\}
The EB-like behavior in the Co$_3$Sn$_2$S$_2$ system is attributed to the coercivity mechanisms. Discrete local magnetic moments serve as the type \uppercase\expandafter{\romannumeral1} nuclei, notably lowering the critical field for magnetic reversal. Tunable EB-like behavior can be realized by simply adjusting the maximum external field applied to the sample, which is actually a control of the hard-magnetic nuclei. An unexpected local magnetic state with strong exchange interaction and higher stability than the majority is suggested in this study. Further investigation is needed to unveil whether this state is linked to the reported spin-orbit polaron. Our research offers new avenues for manipulating the magnetic reversals, thereby controlling the magnetic-topological properties. In addition, the potential local magnetic state can enrich the understanding of the magnetism in the Co$_3$Sn$_2$S$_2$ Weyl system.
\\

\begin{acknowledgments}
The authors thank H. G. Zhang and G. H. Wu for the discussions, and W. Tong and Q. Li for the technological support in the electric measurements.
\end{acknowledgments}
\bigskip

\bigskip

\bibliography{CSSHc}

\begin{thebibliography}{56}%
\makeatletter
\providecommand \@ifxundefined [1]{%
 \@ifx{#1\undefined}
}%
\providecommand \@ifnum [1]{%
 \ifnum #1\expandafter \@firstoftwo
 \else \expandafter \@secondoftwo
 \fi
}%
\providecommand \@ifx [1]{%
 \ifx #1\expandafter \@firstoftwo
 \else \expandafter \@secondoftwo
 \fi
}%
\providecommand \natexlab [1]{#1}%
\providecommand \enquote  [1]{``#1''}%
\providecommand \bibnamefont  [1]{#1}%
\providecommand \bibfnamefont [1]{#1}%
\providecommand \citenamefont [1]{#1}%
\providecommand \href@noop [0]{\@secondoftwo}%
\providecommand \href [0]{\begingroup \@sanitize@url \@href}%
\providecommand \@href[1]{\@@startlink{#1}\@@href}%
\providecommand \@@href[1]{\endgroup#1\@@endlink}%
\providecommand \@sanitize@url [0]{\catcode `\\12\catcode `\$12\catcode `\&12\catcode `\#12\catcode `\^12\catcode `\_12\catcode `\%12\relax}%
\providecommand \@@startlink[1]{}%
\providecommand \@@endlink[0]{}%
\providecommand \url  [0]{\begingroup\@sanitize@url \@url }%
\providecommand \@url [1]{\endgroup\@href {#1}{\urlprefix }}%
\providecommand \urlprefix  [0]{URL }%
\providecommand \Eprint [0]{\href }%
\providecommand \doibase [0]{http://dx.doi.org/}%
\providecommand \selectlanguage [0]{\@gobble}%
\providecommand \bibinfo  [0]{\@secondoftwo}%
\providecommand \bibfield  [0]{\@secondoftwo}%
\providecommand \translation [1]{[#1]}%
\providecommand \BibitemOpen [0]{}%
\providecommand \bibitemStop [0]{}%
\providecommand \bibitemNoStop [0]{.\EOS\space}%
\providecommand \EOS [0]{\spacefactor3000\relax}%
\providecommand \BibitemShut  [1]{\csname bibitem#1\endcsname}%
\let\auto@bib@innerbib\@empty
\bibitem [{\citenamefont {Belopolski}\ \emph {et~al.}(2019)\citenamefont {Belopolski}, \citenamefont {Manna}, \citenamefont {Sanchez}, \citenamefont {Chang}, \citenamefont {Ernst}, \citenamefont {Yin}, \citenamefont {Zhang}, \citenamefont {Cochran}, \citenamefont {Shumiya}, \citenamefont {Zheng}, \citenamefont {Singh}, \citenamefont {Bian}, \citenamefont {Multer}, \citenamefont {Litskevich}, \citenamefont {Zhou}, \citenamefont {Huang}, \citenamefont {Wang}, \citenamefont {Chang}, \citenamefont {Xu}, \citenamefont {Bansil}, \citenamefont {Felser}, \citenamefont {Lin},\ and\ \citenamefont {Hasan}}]{RN830}%
  \BibitemOpen
  \bibfield  {author} {\bibinfo {author} {\bibfnamefont {I.}~\bibnamefont {Belopolski}}, \bibinfo {author} {\bibfnamefont {K.}~\bibnamefont {Manna}}, \bibinfo {author} {\bibfnamefont {D.~S.}\ \bibnamefont {Sanchez}}, \bibinfo {author} {\bibfnamefont {G.}~\bibnamefont {Chang}}, \bibinfo {author} {\bibfnamefont {B.}~\bibnamefont {Ernst}}, \bibinfo {author} {\bibfnamefont {J.}~\bibnamefont {Yin}}, \bibinfo {author} {\bibfnamefont {S.~S.}\ \bibnamefont {Zhang}}, \bibinfo {author} {\bibfnamefont {T.}~\bibnamefont {Cochran}}, \bibinfo {author} {\bibfnamefont {N.}~\bibnamefont {Shumiya}}, \bibinfo {author} {\bibfnamefont {H.}~\bibnamefont {Zheng}}, \bibinfo {author} {\bibfnamefont {B.}~\bibnamefont {Singh}}, \bibinfo {author} {\bibfnamefont {G.}~\bibnamefont {Bian}}, \bibinfo {author} {\bibfnamefont {D.}~\bibnamefont {Multer}}, \bibinfo {author} {\bibfnamefont {M.}~\bibnamefont {Litskevich}}, \bibinfo {author} {\bibfnamefont {X.}~\bibnamefont {Zhou}}, \bibinfo {author} {\bibfnamefont {S.-M.}\ \bibnamefont {Huang}}, \bibinfo {author} {\bibfnamefont {B.}~\bibnamefont {Wang}}, \bibinfo {author} {\bibfnamefont {T.-R.}\ \bibnamefont {Chang}}, \bibinfo {author} {\bibfnamefont {S.-Y.}\ \bibnamefont {Xu}}, \bibinfo {author} {\bibfnamefont {A.}~\bibnamefont {Bansil}}, \bibinfo {author} {\bibfnamefont {C.}~\bibnamefont {Felser}}, \bibinfo {author} {\bibfnamefont {H.}~\bibnamefont {Lin}}, \ and\ \bibinfo {author} {\bibfnamefont {M.~Z.}\ \bibnamefont {Hasan}},\ }\bibfield  {title} {\enquote {\bibinfo {title} {Discovery of topological weyl fermion lines and drumhead surface states in a room temperature magnet},}\ }\href@noop {} {\bibfield  {journal} {\bibinfo  {journal} {Science}\ }\textbf {\bibinfo {volume} {365}},\ \bibinfo {pages} {1278} (\bibinfo {year} {2019})}\BibitemShut {NoStop}%
\bibitem [{\citenamefont {Chang}\ \emph {et~al.}(2016)\citenamefont {Chang}, \citenamefont {Xu}, \citenamefont {Zheng}, \citenamefont {Singh}, \citenamefont {Hsu}, \citenamefont {Bian}, \citenamefont {Alidoust}, \citenamefont {Belopolski}, \citenamefont {Sanchez}, \citenamefont {Zhang}, \citenamefont {Lin},\ and\ \citenamefont {Hasan}}]{RN838}%
  \BibitemOpen
  \bibfield  {author} {\bibinfo {author} {\bibfnamefont {G.}~\bibnamefont {Chang}}, \bibinfo {author} {\bibfnamefont {S.~Y.}\ \bibnamefont {Xu}}, \bibinfo {author} {\bibfnamefont {H.}~\bibnamefont {Zheng}}, \bibinfo {author} {\bibfnamefont {B.}~\bibnamefont {Singh}}, \bibinfo {author} {\bibfnamefont {C.~H.}\ \bibnamefont {Hsu}}, \bibinfo {author} {\bibfnamefont {G.}~\bibnamefont {Bian}}, \bibinfo {author} {\bibfnamefont {N.}~\bibnamefont {Alidoust}}, \bibinfo {author} {\bibfnamefont {I.}~\bibnamefont {Belopolski}}, \bibinfo {author} {\bibfnamefont {D.~S.}\ \bibnamefont {Sanchez}}, \bibinfo {author} {\bibfnamefont {S.}~\bibnamefont {Zhang}}, \bibinfo {author} {\bibfnamefont {H.}~\bibnamefont {Lin}}, \ and\ \bibinfo {author} {\bibfnamefont {M.~Z.}\ \bibnamefont {Hasan}},\ }\bibfield  {title} {\enquote {\bibinfo {title} {Room-temperature magnetic topological weyl fermion and nodal line semimetal states in half-metallic heusler co$_2$tix (x=si, ge, or sn)},}\ }\href@noop {} {\bibfield  {journal} {\bibinfo  {journal} {Sci. Rep.}\ }\textbf {\bibinfo {volume} {6}},\ \bibinfo {pages} {38839} (\bibinfo {year} {2016})}\BibitemShut {NoStop}%
\bibitem [{\citenamefont {Destraz}\ \emph {et~al.}(2020)\citenamefont {Destraz}, \citenamefont {Das}, \citenamefont {Tsirkin}, \citenamefont {Xu}, \citenamefont {Neupert}, \citenamefont {Chang}, \citenamefont {Schilling}, \citenamefont {Grushin}, \citenamefont {Kohlbrecher}, \citenamefont {Keller}, \citenamefont {Puphal}, \citenamefont {Pomjakushina},\ and\ \citenamefont {White}}]{RN837}%
  \BibitemOpen
  \bibfield  {author} {\bibinfo {author} {\bibfnamefont {D.}~\bibnamefont {Destraz}}, \bibinfo {author} {\bibfnamefont {L.}~\bibnamefont {Das}}, \bibinfo {author} {\bibfnamefont {S.~S.}\ \bibnamefont {Tsirkin}}, \bibinfo {author} {\bibfnamefont {Y.}~\bibnamefont {Xu}}, \bibinfo {author} {\bibfnamefont {T.}~\bibnamefont {Neupert}}, \bibinfo {author} {\bibfnamefont {J.}~\bibnamefont {Chang}}, \bibinfo {author} {\bibfnamefont {A.}~\bibnamefont {Schilling}}, \bibinfo {author} {\bibfnamefont {A.~G.}\ \bibnamefont {Grushin}}, \bibinfo {author} {\bibfnamefont {J.}~\bibnamefont {Kohlbrecher}}, \bibinfo {author} {\bibfnamefont {L.}~\bibnamefont {Keller}}, \bibinfo {author} {\bibfnamefont {P.}~\bibnamefont {Puphal}}, \bibinfo {author} {\bibfnamefont {E.}~\bibnamefont {Pomjakushina}}, \ and\ \bibinfo {author} {\bibfnamefont {J.~S.}\ \bibnamefont {White}},\ }\bibfield  {title} {\enquote {\bibinfo {title} {Magnetism and anomalous transport in the weyl semimetal pralge: possible route to axial gauge fields},}\ }\href@noop {} {\bibfield  {journal} {\bibinfo  {journal} {npj Quantum Mater.}\ }\textbf {\bibinfo {volume} {5}},\ \bibinfo {pages} {5} (\bibinfo {year} {2020})}\BibitemShut {NoStop}%
\bibitem [{\citenamefont {Gao}\ \emph {et~al.}(2021)\citenamefont {Gao}, \citenamefont {Xu}, \citenamefont {Li}, \citenamefont {Yi}, \citenamefont {Nie}, \citenamefont {Rao}, \citenamefont {Wang}, \citenamefont {Hu}, \citenamefont {Chen}, \citenamefont {Fan}, \citenamefont {Huang}, \citenamefont {Huang}, \citenamefont {Pryds}, \citenamefont {Shi}, \citenamefont {Wang}, \citenamefont {Shi}, \citenamefont {Xia}, \citenamefont {Qian},\ and\ \citenamefont {Ding}}]{RN1154}%
  \BibitemOpen
  \bibfield  {author} {\bibinfo {author} {\bibfnamefont {S.-Y.}\ \bibnamefont {Gao}}, \bibinfo {author} {\bibfnamefont {S.}~\bibnamefont {Xu}}, \bibinfo {author} {\bibfnamefont {H.}~\bibnamefont {Li}}, \bibinfo {author} {\bibfnamefont {C.-J.}\ \bibnamefont {Yi}}, \bibinfo {author} {\bibfnamefont {S.-M.}\ \bibnamefont {Nie}}, \bibinfo {author} {\bibfnamefont {Z.-C.}\ \bibnamefont {Rao}}, \bibinfo {author} {\bibfnamefont {H.}~\bibnamefont {Wang}}, \bibinfo {author} {\bibfnamefont {Q.-X.}\ \bibnamefont {Hu}}, \bibinfo {author} {\bibfnamefont {X.-Z.}\ \bibnamefont {Chen}}, \bibinfo {author} {\bibfnamefont {W.-H.}\ \bibnamefont {Fan}}, \bibinfo {author} {\bibfnamefont {J.-R.}\ \bibnamefont {Huang}}, \bibinfo {author} {\bibfnamefont {Y.-B.}\ \bibnamefont {Huang}}, \bibinfo {author} {\bibfnamefont {N.}~\bibnamefont {Pryds}}, \bibinfo {author} {\bibfnamefont {M.}~\bibnamefont {Shi}}, \bibinfo {author} {\bibfnamefont {Z.-J.}\ \bibnamefont {Wang}}, \bibinfo {author} {\bibfnamefont {Y.-G.}\ \bibnamefont {Shi}}, \bibinfo {author} {\bibfnamefont {T.-L.}\ \bibnamefont {Xia}}, \bibinfo {author} {\bibfnamefont {T.}~\bibnamefont {Qian}}, \ and\ \bibinfo {author} {\bibfnamefont {H.}~\bibnamefont {Ding}},\ }\bibfield  {title} {\enquote {\bibinfo {title} {Time-reversal symmetry breaking driven topological phase transition in eub$_6$},}\ }\href@noop {} {\bibfield  {journal} {\bibinfo  {journal} {Phys. Rev. X}\ }\textbf {\bibinfo {volume} {11}},\ \bibinfo {pages} {021016} (\bibinfo {year} {2021})}\BibitemShut {NoStop}%
\bibitem [{\citenamefont {Liu}\ \emph {et~al.}(2018)\citenamefont {Liu}, \citenamefont {Sun}, \citenamefont {Kumar}, \citenamefont {Muchler}, \citenamefont {Sun}, \citenamefont {Jiao}, \citenamefont {Yang}, \citenamefont {Liu}, \citenamefont {Liang}, \citenamefont {Xu}, \citenamefont {Kroder}, \citenamefont {Suss}, \citenamefont {Borrmann}, \citenamefont {Shekhar}, \citenamefont {Wang}, \citenamefont {Xi}, \citenamefont {Wang}, \citenamefont {Schnelle}, \citenamefont {Wirth}, \citenamefont {Chen}, \citenamefont {Goennenwein},\ and\ \citenamefont {Felser}}]{RN243}%
  \BibitemOpen
  \bibfield  {author} {\bibinfo {author} {\bibfnamefont {E.~K.}\ \bibnamefont {Liu}}, \bibinfo {author} {\bibfnamefont {Y.}~\bibnamefont {Sun}}, \bibinfo {author} {\bibfnamefont {N.}~\bibnamefont {Kumar}}, \bibinfo {author} {\bibfnamefont {L.}~\bibnamefont {Muchler}}, \bibinfo {author} {\bibfnamefont {A.}~\bibnamefont {Sun}}, \bibinfo {author} {\bibfnamefont {L.}~\bibnamefont {Jiao}}, \bibinfo {author} {\bibfnamefont {S.~Y.}\ \bibnamefont {Yang}}, \bibinfo {author} {\bibfnamefont {D.}~\bibnamefont {Liu}}, \bibinfo {author} {\bibfnamefont {A.}~\bibnamefont {Liang}}, \bibinfo {author} {\bibfnamefont {Q.}~\bibnamefont {Xu}}, \bibinfo {author} {\bibfnamefont {J.}~\bibnamefont {Kroder}}, \bibinfo {author} {\bibfnamefont {V.}~\bibnamefont {Suss}}, \bibinfo {author} {\bibfnamefont {H.}~\bibnamefont {Borrmann}}, \bibinfo {author} {\bibfnamefont {C.}~\bibnamefont {Shekhar}}, \bibinfo {author} {\bibfnamefont {Z.}~\bibnamefont {Wang}}, \bibinfo {author} {\bibfnamefont {C.}~\bibnamefont {Xi}}, \bibinfo {author} {\bibfnamefont {W.}~\bibnamefont {Wang}}, \bibinfo {author} {\bibfnamefont {W.}~\bibnamefont {Schnelle}}, \bibinfo {author} {\bibfnamefont {S.}~\bibnamefont {Wirth}}, \bibinfo {author} {\bibfnamefont {Y.}~\bibnamefont {Chen}}, \bibinfo {author} {\bibfnamefont {S.~T.~B.}\ \bibnamefont {Goennenwein}}, \ and\ \bibinfo {author} {\bibfnamefont {C.}~\bibnamefont {Felser}},\ }\bibfield  {title} {\enquote {\bibinfo {title} {Giant anomalous hall effect in a ferromagnetic kagome-lattice semimetal},}\ }\href@noop {} {\bibfield  {journal} {\bibinfo  {journal} {Nat. Phys.}\ }\textbf {\bibinfo {volume} {14}},\ \bibinfo {pages} {1125} (\bibinfo {year} {2018})}\BibitemShut {NoStop}%
\bibitem [{\citenamefont {Nie}\ \emph {et~al.}(2020)\citenamefont {Nie}, \citenamefont {Sun}, \citenamefont {Prinz}, \citenamefont {Wang}, \citenamefont {Weng}, \citenamefont {Fang},\ and\ \citenamefont {Dai}}]{RN1178}%
  \BibitemOpen
  \bibfield  {author} {\bibinfo {author} {\bibfnamefont {S.}~\bibnamefont {Nie}}, \bibinfo {author} {\bibfnamefont {Y.}~\bibnamefont {Sun}}, \bibinfo {author} {\bibfnamefont {F.~B.}\ \bibnamefont {Prinz}}, \bibinfo {author} {\bibfnamefont {Z.}~\bibnamefont {Wang}}, \bibinfo {author} {\bibfnamefont {H.}~\bibnamefont {Weng}}, \bibinfo {author} {\bibfnamefont {Z.}~\bibnamefont {Fang}}, \ and\ \bibinfo {author} {\bibfnamefont {X.}~\bibnamefont {Dai}},\ }\bibfield  {title} {\enquote {\bibinfo {title} {Magnetic semimetals and quantized anomalous hall effect in eub$_6$},}\ }\href@noop {} {\bibfield  {journal} {\bibinfo  {journal} {Phys. Rev. Lett.}\ }\textbf {\bibinfo {volume} {124}},\ \bibinfo {pages} {076403} (\bibinfo {year} {2020})}\BibitemShut {NoStop}%
\bibitem [{\citenamefont {Tsai}\ \emph {et~al.}(2020)\citenamefont {Tsai}, \citenamefont {Higo}, \citenamefont {Kondou}, \citenamefont {Nomoto}, \citenamefont {Sakai}, \citenamefont {Kobayashi}, \citenamefont {Nakano}, \citenamefont {Yakushiji}, \citenamefont {Arita}, \citenamefont {Miwa}, \citenamefont {Otani},\ and\ \citenamefont {Nakatsuji}}]{RN1447}%
  \BibitemOpen
  \bibfield  {author} {\bibinfo {author} {\bibfnamefont {H.}~\bibnamefont {Tsai}}, \bibinfo {author} {\bibfnamefont {T.}~\bibnamefont {Higo}}, \bibinfo {author} {\bibfnamefont {K.}~\bibnamefont {Kondou}}, \bibinfo {author} {\bibfnamefont {T.}~\bibnamefont {Nomoto}}, \bibinfo {author} {\bibfnamefont {A.}~\bibnamefont {Sakai}}, \bibinfo {author} {\bibfnamefont {A.}~\bibnamefont {Kobayashi}}, \bibinfo {author} {\bibfnamefont {T.}~\bibnamefont {Nakano}}, \bibinfo {author} {\bibfnamefont {K.}~\bibnamefont {Yakushiji}}, \bibinfo {author} {\bibfnamefont {R.}~\bibnamefont {Arita}}, \bibinfo {author} {\bibfnamefont {S.}~\bibnamefont {Miwa}}, \bibinfo {author} {\bibfnamefont {Y.}~\bibnamefont {Otani}}, \ and\ \bibinfo {author} {\bibfnamefont {S.}~\bibnamefont {Nakatsuji}},\ }\bibfield  {title} {\enquote {\bibinfo {title} {Electrical manipulation of a topological antiferromagnetic state},}\ }\href@noop {} {\bibfield  {journal} {\bibinfo  {journal} {Nature}\ }\textbf {\bibinfo {volume} {580}},\ \bibinfo {pages} {608} (\bibinfo {year} {2020})}\BibitemShut {NoStop}%
\bibitem [{\citenamefont {Wan}\ \emph {et~al.}(2011)\citenamefont {Wan}, \citenamefont {Turner}, \citenamefont {Vishwanath},\ and\ \citenamefont {Savrasov}}]{RN842}%
  \BibitemOpen
  \bibfield  {author} {\bibinfo {author} {\bibfnamefont {X.}~\bibnamefont {Wan}}, \bibinfo {author} {\bibfnamefont {A.~M.}\ \bibnamefont {Turner}}, \bibinfo {author} {\bibfnamefont {A.}~\bibnamefont {Vishwanath}}, \ and\ \bibinfo {author} {\bibfnamefont {S.~Y.}\ \bibnamefont {Savrasov}},\ }\bibfield  {title} {\enquote {\bibinfo {title} {Topological semimetal and fermi-arc surface states in the electronic structure of pyrochlore iridates},}\ }\href@noop {} {\bibfield  {journal} {\bibinfo  {journal} {Phys. Rev. B}\ }\textbf {\bibinfo {volume} {83}},\ \bibinfo {pages} {205101} (\bibinfo {year} {2011})}\BibitemShut {NoStop}%
\bibitem [{\citenamefont {Wang}\ \emph {et~al.}(2018)\citenamefont {Wang}, \citenamefont {Xu}, \citenamefont {Lou}, \citenamefont {Liu}, \citenamefont {Li}, \citenamefont {Huang}, \citenamefont {Shen}, \citenamefont {Weng}, \citenamefont {Wang},\ and\ \citenamefont {Lei}}]{RN587}%
  \BibitemOpen
  \bibfield  {author} {\bibinfo {author} {\bibfnamefont {Q.}~\bibnamefont {Wang}}, \bibinfo {author} {\bibfnamefont {Y.}~\bibnamefont {Xu}}, \bibinfo {author} {\bibfnamefont {R.}~\bibnamefont {Lou}}, \bibinfo {author} {\bibfnamefont {Z.}~\bibnamefont {Liu}}, \bibinfo {author} {\bibfnamefont {M.}~\bibnamefont {Li}}, \bibinfo {author} {\bibfnamefont {Y.}~\bibnamefont {Huang}}, \bibinfo {author} {\bibfnamefont {D.}~\bibnamefont {Shen}}, \bibinfo {author} {\bibfnamefont {H.}~\bibnamefont {Weng}}, \bibinfo {author} {\bibfnamefont {S.}~\bibnamefont {Wang}}, \ and\ \bibinfo {author} {\bibfnamefont {H.}~\bibnamefont {Lei}},\ }\bibfield  {title} {\enquote {\bibinfo {title} {Large intrinsic anomalous hall effect in half-metallic ferromagnet co$_3$sn$_2$s$_2$ with magnetic weyl fermions},}\ }\href@noop {} {\bibfield  {journal} {\bibinfo  {journal} {Nat. Commun.}\ }\textbf {\bibinfo {volume} {9}},\ \bibinfo {pages} {3681} (\bibinfo {year} {2018})}\BibitemShut {NoStop}%
\bibitem [{\citenamefont {Wang}\ \emph {et~al.}(2016)\citenamefont {Wang}, \citenamefont {Vergniory}, \citenamefont {Kushwaha}, \citenamefont {Hirschberger}, \citenamefont {Chulkov}, \citenamefont {Ernst}, \citenamefont {Ong}, \citenamefont {Cava},\ and\ \citenamefont {Bernevig}}]{RN839}%
  \BibitemOpen
  \bibfield  {author} {\bibinfo {author} {\bibfnamefont {Z.}~\bibnamefont {Wang}}, \bibinfo {author} {\bibfnamefont {M.~G.}\ \bibnamefont {Vergniory}}, \bibinfo {author} {\bibfnamefont {S.}~\bibnamefont {Kushwaha}}, \bibinfo {author} {\bibfnamefont {M.}~\bibnamefont {Hirschberger}}, \bibinfo {author} {\bibfnamefont {E.~V.}\ \bibnamefont {Chulkov}}, \bibinfo {author} {\bibfnamefont {A.}~\bibnamefont {Ernst}}, \bibinfo {author} {\bibfnamefont {N.~P.}\ \bibnamefont {Ong}}, \bibinfo {author} {\bibfnamefont {R.~J.}\ \bibnamefont {Cava}}, \ and\ \bibinfo {author} {\bibfnamefont {B.~A.}\ \bibnamefont {Bernevig}},\ }\bibfield  {title} {\enquote {\bibinfo {title} {Time-reversal-breaking weyl fermions in magnetic heusler alloys},}\ }\href@noop {} {\bibfield  {journal} {\bibinfo  {journal} {Phys. Rev. Lett.}\ }\textbf {\bibinfo {volume} {117}},\ \bibinfo {pages} {236401} (\bibinfo {year} {2016})}\BibitemShut {NoStop}%
\bibitem [{\citenamefont {Xu}\ \emph {et~al.}(2011)\citenamefont {Xu}, \citenamefont {Weng}, \citenamefont {Wang}, \citenamefont {Dai},\ and\ \citenamefont {Fang}}]{RN841}%
  \BibitemOpen
  \bibfield  {author} {\bibinfo {author} {\bibfnamefont {G.}~\bibnamefont {Xu}}, \bibinfo {author} {\bibfnamefont {H.}~\bibnamefont {Weng}}, \bibinfo {author} {\bibfnamefont {Z.}~\bibnamefont {Wang}}, \bibinfo {author} {\bibfnamefont {X.}~\bibnamefont {Dai}}, \ and\ \bibinfo {author} {\bibfnamefont {Z.}~\bibnamefont {Fang}},\ }\bibfield  {title} {\enquote {\bibinfo {title} {Chern semimetal and the quantized anomalous hall effect in hgcr$_2$se$_4$},}\ }\href@noop {} {\bibfield  {journal} {\bibinfo  {journal} {Phys. Rev. Lett.}\ }\textbf {\bibinfo {volume} {107}},\ \bibinfo {pages} {186806} (\bibinfo {year} {2011})}\BibitemShut {NoStop}%
\bibitem [{\citenamefont {Yin}\ \emph {et~al.}(2019)\citenamefont {Yin}, \citenamefont {Zhang}, \citenamefont {Chang}, \citenamefont {Wang}, \citenamefont {Tsirkin}, \citenamefont {Guguchia}, \citenamefont {Lian}, \citenamefont {Zhou}, \citenamefont {Jiang}, \citenamefont {Belopolski}, \citenamefont {Shumiya}, \citenamefont {Multer}, \citenamefont {Litskevich}, \citenamefont {Cochran}, \citenamefont {Lin}, \citenamefont {Wang}, \citenamefont {Neupert}, \citenamefont {Jia}, \citenamefont {Lei},\ and\ \citenamefont {Hasan}}]{RN591}%
  \BibitemOpen
  \bibfield  {author} {\bibinfo {author} {\bibfnamefont {J.-X.}\ \bibnamefont {Yin}}, \bibinfo {author} {\bibfnamefont {S.~S.}\ \bibnamefont {Zhang}}, \bibinfo {author} {\bibfnamefont {G.}~\bibnamefont {Chang}}, \bibinfo {author} {\bibfnamefont {Q.}~\bibnamefont {Wang}}, \bibinfo {author} {\bibfnamefont {S.~S.}\ \bibnamefont {Tsirkin}}, \bibinfo {author} {\bibfnamefont {Z.}~\bibnamefont {Guguchia}}, \bibinfo {author} {\bibfnamefont {B.}~\bibnamefont {Lian}}, \bibinfo {author} {\bibfnamefont {H.}~\bibnamefont {Zhou}}, \bibinfo {author} {\bibfnamefont {K.}~\bibnamefont {Jiang}}, \bibinfo {author} {\bibfnamefont {I.}~\bibnamefont {Belopolski}}, \bibinfo {author} {\bibfnamefont {N.}~\bibnamefont {Shumiya}}, \bibinfo {author} {\bibfnamefont {D.}~\bibnamefont {Multer}}, \bibinfo {author} {\bibfnamefont {M.}~\bibnamefont {Litskevich}}, \bibinfo {author} {\bibfnamefont {T.~A.}\ \bibnamefont {Cochran}}, \bibinfo {author} {\bibfnamefont {H.}~\bibnamefont {Lin}}, \bibinfo {author} {\bibfnamefont {Z.}~\bibnamefont {Wang}}, \bibinfo {author} {\bibfnamefont {T.}~\bibnamefont {Neupert}}, \bibinfo {author} {\bibfnamefont {S.}~\bibnamefont {Jia}}, \bibinfo {author} {\bibfnamefont {H.}~\bibnamefont {Lei}}, \ and\ \bibinfo {author} {\bibfnamefont {M.~Z.}\ \bibnamefont {Hasan}},\ }\bibfield  {title} {\enquote {\bibinfo {title} {Negative flat band magnetism in a spin–orbit-coupled correlated kagome magnet},}\ }\href@noop {} {\bibfield  {journal} {\bibinfo  {journal} {Nat. Phys.}\ }\textbf {\bibinfo {volume} {15}},\ \bibinfo {pages} {443} (\bibinfo {year} {2019})}\BibitemShut {NoStop}%
\bibitem [{\citenamefont {Liu}\ \emph {et~al.}(2019)\citenamefont {Liu}, \citenamefont {Liang}, \citenamefont {Liu}, \citenamefont {Xu}, \citenamefont {Li}, \citenamefont {Chen}, \citenamefont {Pei}, \citenamefont {Shi}, \citenamefont {Mo}, \citenamefont {Dudin}, \citenamefont {Kim}, \citenamefont {Cacho}, \citenamefont {Li}, \citenamefont {Sun}, \citenamefont {Yang}, \citenamefont {Liu}, \citenamefont {Parkin}, \citenamefont {Felser},\ and\ \citenamefont {Chen}}]{RN504}%
  \BibitemOpen
  \bibfield  {author} {\bibinfo {author} {\bibfnamefont {D.~F.}\ \bibnamefont {Liu}}, \bibinfo {author} {\bibfnamefont {A.~J.}\ \bibnamefont {Liang}}, \bibinfo {author} {\bibfnamefont {E.~K.}\ \bibnamefont {Liu}}, \bibinfo {author} {\bibfnamefont {Q.~N.}\ \bibnamefont {Xu}}, \bibinfo {author} {\bibfnamefont {Y.~W.}\ \bibnamefont {Li}}, \bibinfo {author} {\bibfnamefont {C.}~\bibnamefont {Chen}}, \bibinfo {author} {\bibfnamefont {D.}~\bibnamefont {Pei}}, \bibinfo {author} {\bibfnamefont {W.~J.}\ \bibnamefont {Shi}}, \bibinfo {author} {\bibfnamefont {S.~K.}\ \bibnamefont {Mo}}, \bibinfo {author} {\bibfnamefont {P.}~\bibnamefont {Dudin}}, \bibinfo {author} {\bibfnamefont {T.}~\bibnamefont {Kim}}, \bibinfo {author} {\bibfnamefont {C.}~\bibnamefont {Cacho}}, \bibinfo {author} {\bibfnamefont {G.}~\bibnamefont {Li}}, \bibinfo {author} {\bibfnamefont {Y.}~\bibnamefont {Sun}}, \bibinfo {author} {\bibfnamefont {L.~X.}\ \bibnamefont {Yang}}, \bibinfo {author} {\bibfnamefont {Z.~K.}\ \bibnamefont {Liu}}, \bibinfo {author} {\bibfnamefont {S.~S.~P.}\ \bibnamefont {Parkin}}, \bibinfo {author} {\bibfnamefont {C.}~\bibnamefont {Felser}}, \ and\ \bibinfo {author} {\bibfnamefont {Y.~L.}\ \bibnamefont {Chen}},\ }\bibfield  {title} {\enquote {\bibinfo {title} {Magnetic weyl semimetal phase in a kagomé crystal},}\ }\href@noop {} {\bibfield  {journal} {\bibinfo  {journal} {Science}\ }\textbf {\bibinfo {volume} {365}},\ \bibinfo {pages} {1282} (\bibinfo {year} {2019})}\BibitemShut {NoStop}%
\bibitem [{\citenamefont {Morali}\ \emph {et~al.}(2019)\citenamefont {Morali}, \citenamefont {Batabyal}, \citenamefont {Nag}, \citenamefont {Liu}, \citenamefont {Xu}, \citenamefont {Sun}, \citenamefont {Yan}, \citenamefont {Felser}, \citenamefont {Avraham},\ and\ \citenamefont {Beidenkopf}}]{RN505}%
  \BibitemOpen
  \bibfield  {author} {\bibinfo {author} {\bibfnamefont {N.}~\bibnamefont {Morali}}, \bibinfo {author} {\bibfnamefont {R.}~\bibnamefont {Batabyal}}, \bibinfo {author} {\bibfnamefont {P.~K.}\ \bibnamefont {Nag}}, \bibinfo {author} {\bibfnamefont {E.}~\bibnamefont {Liu}}, \bibinfo {author} {\bibfnamefont {Q.}~\bibnamefont {Xu}}, \bibinfo {author} {\bibfnamefont {Y.}~\bibnamefont {Sun}}, \bibinfo {author} {\bibfnamefont {B.}~\bibnamefont {Yan}}, \bibinfo {author} {\bibfnamefont {C.}~\bibnamefont {Felser}}, \bibinfo {author} {\bibfnamefont {N.}~\bibnamefont {Avraham}}, \ and\ \bibinfo {author} {\bibfnamefont {H.}~\bibnamefont {Beidenkopf}},\ }\bibfield  {title} {\enquote {\bibinfo {title} {Fermi-arc diversity on surface terminations of the magnetic weyl semimetal co$_3$sn$_2$s$_2$},}\ }\href@noop {} {\bibfield  {journal} {\bibinfo  {journal} {Science}\ }\textbf {\bibinfo {volume} {365}},\ \bibinfo {pages} {1286} (\bibinfo {year} {2019})}\BibitemShut {NoStop}%
\bibitem [{\citenamefont {Schnelle}\ \emph {et~al.}(2013)\citenamefont {Schnelle}, \citenamefont {Leithe-Jasper}, \citenamefont {Rosner}, \citenamefont {Schappacher}, \citenamefont {Pöttgen}, \citenamefont {Pielnhofer},\ and\ \citenamefont {Weihrich}}]{RN536}%
  \BibitemOpen
  \bibfield  {author} {\bibinfo {author} {\bibfnamefont {W.}~\bibnamefont {Schnelle}}, \bibinfo {author} {\bibfnamefont {A.}~\bibnamefont {Leithe-Jasper}}, \bibinfo {author} {\bibfnamefont {H.}~\bibnamefont {Rosner}}, \bibinfo {author} {\bibfnamefont {F.~M.}\ \bibnamefont {Schappacher}}, \bibinfo {author} {\bibfnamefont {R.}~\bibnamefont {Pöttgen}}, \bibinfo {author} {\bibfnamefont {F.}~\bibnamefont {Pielnhofer}}, \ and\ \bibinfo {author} {\bibfnamefont {R.}~\bibnamefont {Weihrich}},\ }\bibfield  {title} {\enquote {\bibinfo {title} {Ferromagnetic ordering and half-metallic state of sn$_2$co$_3$s$_2$ with the shandite-type structure},}\ }\href@noop {} {\bibfield  {journal} {\bibinfo  {journal} {Phys. Rev. B}\ }\textbf {\bibinfo {volume} {88}},\ \bibinfo {pages} {144404} (\bibinfo {year} {2013})}\BibitemShut {NoStop}%
\bibitem [{\citenamefont {Vaqueiro}\ and\ \citenamefont {Sobany}(2009)}]{RN535}%
  \BibitemOpen
  \bibfield  {author} {\bibinfo {author} {\bibfnamefont {P.}~\bibnamefont {Vaqueiro}}\ and\ \bibinfo {author} {\bibfnamefont {G.~G.}\ \bibnamefont {Sobany}},\ }\bibfield  {title} {\enquote {\bibinfo {title} {A powder neutron diffraction study of the metallic ferromagnet co$_3$sn$_2$s$_2$},}\ }\href@noop {} {\bibfield  {journal} {\bibinfo  {journal} {Solid State Sci.}\ }\textbf {\bibinfo {volume} {11}},\ \bibinfo {pages} {513} (\bibinfo {year} {2009})}\BibitemShut {NoStop}%
\bibitem [{\citenamefont {Weihrich}, \citenamefont {Anusca},\ and\ \citenamefont {Zabel}(2005)}]{RN537}%
  \BibitemOpen
  \bibfield  {author} {\bibinfo {author} {\bibfnamefont {R.}~\bibnamefont {Weihrich}}, \bibinfo {author} {\bibfnamefont {I.}~\bibnamefont {Anusca}}, \ and\ \bibinfo {author} {\bibfnamefont {M.}~\bibnamefont {Zabel}},\ }\bibfield  {title} {\enquote {\bibinfo {title} {Half-antiperovskites: Structure and type-antitype relations of shandites m$_3/2$as (m = co, ni; a = in, sn)},}\ }\href@noop {} {\bibfield  {journal} {\bibinfo  {journal} {Z. Anorg. Allg. Chem.}\ }\textbf {\bibinfo {volume} {631}},\ \bibinfo {pages} {1463} (\bibinfo {year} {2005})}\BibitemShut {NoStop}%
\bibitem [{\citenamefont {Guin}\ \emph {et~al.}(2019{\natexlab{a}})\citenamefont {Guin}, \citenamefont {Vir}, \citenamefont {Zhang}, \citenamefont {Kumar}, \citenamefont {Watzman}, \citenamefont {Fu}, \citenamefont {Liu}, \citenamefont {Manna}, \citenamefont {Schnelle}, \citenamefont {Gooth}, \citenamefont {Shekhar}, \citenamefont {Sun},\ and\ \citenamefont {Felser}}]{RN1225}%
  \BibitemOpen
  \bibfield  {author} {\bibinfo {author} {\bibfnamefont {S.~N.}\ \bibnamefont {Guin}}, \bibinfo {author} {\bibfnamefont {P.}~\bibnamefont {Vir}}, \bibinfo {author} {\bibfnamefont {Y.}~\bibnamefont {Zhang}}, \bibinfo {author} {\bibfnamefont {N.}~\bibnamefont {Kumar}}, \bibinfo {author} {\bibfnamefont {S.~J.}\ \bibnamefont {Watzman}}, \bibinfo {author} {\bibfnamefont {C.}~\bibnamefont {Fu}}, \bibinfo {author} {\bibfnamefont {E.}~\bibnamefont {Liu}}, \bibinfo {author} {\bibfnamefont {K.}~\bibnamefont {Manna}}, \bibinfo {author} {\bibfnamefont {W.}~\bibnamefont {Schnelle}}, \bibinfo {author} {\bibfnamefont {J.}~\bibnamefont {Gooth}}, \bibinfo {author} {\bibfnamefont {C.}~\bibnamefont {Shekhar}}, \bibinfo {author} {\bibfnamefont {Y.}~\bibnamefont {Sun}}, \ and\ \bibinfo {author} {\bibfnamefont {C.}~\bibnamefont {Felser}},\ }\bibfield  {title} {\enquote {\bibinfo {title} {Zero-field nernst effect in a ferromagnetic kagome-lattice weyl-semimetal co$_3$sn$_2$s$_2$},}\ }\href@noop {} {\bibfield  {journal} {\bibinfo  {journal} {Adv. Mater.}\ }\textbf {\bibinfo {volume} {31}},\ \bibinfo {pages} {e1806622} (\bibinfo {year} {2019}{\natexlab{a}})}\BibitemShut {NoStop}%
\bibitem [{\citenamefont {Jiang}\ \emph {et~al.}(2022)\citenamefont {Jiang}, \citenamefont {Zhao}, \citenamefont {Qian}, \citenamefont {Zhang}, \citenamefont {Qiang}, \citenamefont {Wang}, \citenamefont {Bi}, \citenamefont {Fan}, \citenamefont {Lu}, \citenamefont {Liu},\ and\ \citenamefont {Wu}}]{RN1490}%
  \BibitemOpen
  \bibfield  {author} {\bibinfo {author} {\bibfnamefont {B.}~\bibnamefont {Jiang}}, \bibinfo {author} {\bibfnamefont {J.}~\bibnamefont {Zhao}}, \bibinfo {author} {\bibfnamefont {J.}~\bibnamefont {Qian}}, \bibinfo {author} {\bibfnamefont {S.}~\bibnamefont {Zhang}}, \bibinfo {author} {\bibfnamefont {X.}~\bibnamefont {Qiang}}, \bibinfo {author} {\bibfnamefont {L.}~\bibnamefont {Wang}}, \bibinfo {author} {\bibfnamefont {R.}~\bibnamefont {Bi}}, \bibinfo {author} {\bibfnamefont {J.}~\bibnamefont {Fan}}, \bibinfo {author} {\bibfnamefont {H.~Z.}\ \bibnamefont {Lu}}, \bibinfo {author} {\bibfnamefont {E.}~\bibnamefont {Liu}}, \ and\ \bibinfo {author} {\bibfnamefont {X.}~\bibnamefont {Wu}},\ }\bibfield  {title} {\enquote {\bibinfo {title} {Antisymmetric seebeck effect in a tilted weyl semimetal},}\ }\href@noop {} {\bibfield  {journal} {\bibinfo  {journal} {Phys. Rev. Lett.}\ }\textbf {\bibinfo {volume} {129}},\ \bibinfo {pages} {056601} (\bibinfo {year} {2022})}\BibitemShut {NoStop}%
\bibitem [{\citenamefont {Noguchi}\ \emph {et~al.}(2024)\citenamefont {Noguchi}, \citenamefont {Fujiwara}, \citenamefont {Yanagi}, \citenamefont {Suzuki}, \citenamefont {Hirai}, \citenamefont {Seki}, \citenamefont {Uchida},\ and\ \citenamefont {Tsukazaki}}]{RN1500}%
  \BibitemOpen
  \bibfield  {author} {\bibinfo {author} {\bibfnamefont {S.}~\bibnamefont {Noguchi}}, \bibinfo {author} {\bibfnamefont {K.}~\bibnamefont {Fujiwara}}, \bibinfo {author} {\bibfnamefont {Y.}~\bibnamefont {Yanagi}}, \bibinfo {author} {\bibfnamefont {M.-T.}\ \bibnamefont {Suzuki}}, \bibinfo {author} {\bibfnamefont {T.}~\bibnamefont {Hirai}}, \bibinfo {author} {\bibfnamefont {T.}~\bibnamefont {Seki}}, \bibinfo {author} {\bibfnamefont {K.-i.}\ \bibnamefont {Uchida}}, \ and\ \bibinfo {author} {\bibfnamefont {A.}~\bibnamefont {Tsukazaki}},\ }\bibfield  {title} {\enquote {\bibinfo {title} {Bipolarity of large anomalous nernst effect in weyl magnet-based alloy films},}\ }\href@noop {} {\bibfield  {journal} {\bibinfo  {journal} {Nat. Phys.}\ }\textbf {\bibinfo {volume} {20}},\ \bibinfo {pages} {254} (\bibinfo {year} {2024})}\BibitemShut {NoStop}%
\bibitem [{\citenamefont {Pan}\ \emph {et~al.}(2024)\citenamefont {Pan}, \citenamefont {He}, \citenamefont {Wang},\ and\ \citenamefont {Felser}}]{RN1498}%
  \BibitemOpen
  \bibfield  {author} {\bibinfo {author} {\bibfnamefont {Y.}~\bibnamefont {Pan}}, \bibinfo {author} {\bibfnamefont {B.}~\bibnamefont {He}}, \bibinfo {author} {\bibfnamefont {H.}~\bibnamefont {Wang}}, \ and\ \bibinfo {author} {\bibfnamefont {C.}~\bibnamefont {Felser}},\ }\bibfield  {title} {\enquote {\bibinfo {title} {Topological materials for high performance transverse thermoelectrics},}\ }\href@noop {} {\bibfield  {journal} {\bibinfo  {journal} {Next Energy}\ }\textbf {\bibinfo {volume} {2}},\ \bibinfo {pages} {100103} (\bibinfo {year} {2024})}\BibitemShut {NoStop}%
\bibitem [{\citenamefont {Lau}\ \emph {et~al.}(2023)\citenamefont {Lau}, \citenamefont {Ikeda}, \citenamefont {Fujiwara}, \citenamefont {Ozawa}, \citenamefont {Zheng}, \citenamefont {Seki}, \citenamefont {Nomura}, \citenamefont {Du}, \citenamefont {Wu}, \citenamefont {Tsukazaki},\ and\ \citenamefont {Takanashi}}]{RN1548}%
  \BibitemOpen
  \bibfield  {author} {\bibinfo {author} {\bibfnamefont {Y.-C.}\ \bibnamefont {Lau}}, \bibinfo {author} {\bibfnamefont {J.}~\bibnamefont {Ikeda}}, \bibinfo {author} {\bibfnamefont {K.}~\bibnamefont {Fujiwara}}, \bibinfo {author} {\bibfnamefont {A.}~\bibnamefont {Ozawa}}, \bibinfo {author} {\bibfnamefont {J.}~\bibnamefont {Zheng}}, \bibinfo {author} {\bibfnamefont {T.}~\bibnamefont {Seki}}, \bibinfo {author} {\bibfnamefont {K.}~\bibnamefont {Nomura}}, \bibinfo {author} {\bibfnamefont {L.}~\bibnamefont {Du}}, \bibinfo {author} {\bibfnamefont {Q.}~\bibnamefont {Wu}}, \bibinfo {author} {\bibfnamefont {A.}~\bibnamefont {Tsukazaki}}, \ and\ \bibinfo {author} {\bibfnamefont {K.}~\bibnamefont {Takanashi}},\ }\bibfield  {title} {\enquote {\bibinfo {title} {Intercorrelated anomalous hall and spin hall effect in kagome-lattice co$_3$sn$_2$s$_2$-based shandite films},}\ }\href@noop {} {\bibfield  {journal} {\bibinfo  {journal} {Phys. Rev. B}\ }\textbf {\bibinfo {volume} {108}},\ \bibinfo {pages} {064429} (\bibinfo {year} {2023})}\BibitemShut {NoStop}%
\bibitem [{\citenamefont {Wang}\ \emph {et~al.}(2022)\citenamefont {Wang}, \citenamefont {Zeng}, \citenamefont {Yuan}, \citenamefont {Zeng}, \citenamefont {Gu}, \citenamefont {Xu}, \citenamefont {Wang}, \citenamefont {Han}, \citenamefont {Nomura}, \citenamefont {Wang}, \citenamefont {Liu}, \citenamefont {Hou},\ and\ \citenamefont {Ye}}]{RN1443}%
  \BibitemOpen
  \bibfield  {author} {\bibinfo {author} {\bibfnamefont {Q.}~\bibnamefont {Wang}}, \bibinfo {author} {\bibfnamefont {Y.}~\bibnamefont {Zeng}}, \bibinfo {author} {\bibfnamefont {K.}~\bibnamefont {Yuan}}, \bibinfo {author} {\bibfnamefont {Q.}~\bibnamefont {Zeng}}, \bibinfo {author} {\bibfnamefont {P.}~\bibnamefont {Gu}}, \bibinfo {author} {\bibfnamefont {X.}~\bibnamefont {Xu}}, \bibinfo {author} {\bibfnamefont {H.}~\bibnamefont {Wang}}, \bibinfo {author} {\bibfnamefont {Z.}~\bibnamefont {Han}}, \bibinfo {author} {\bibfnamefont {K.}~\bibnamefont {Nomura}}, \bibinfo {author} {\bibfnamefont {W.}~\bibnamefont {Wang}}, \bibinfo {author} {\bibfnamefont {E.}~\bibnamefont {Liu}}, \bibinfo {author} {\bibfnamefont {Y.}~\bibnamefont {Hou}}, \ and\ \bibinfo {author} {\bibfnamefont {Y.}~\bibnamefont {Ye}},\ }\bibfield  {title} {\enquote {\bibinfo {title} {Magnetism modulation in co$_3$sn$_2$s$_2$ by current-assisted domain wall motion},}\ }\href@noop {} {\bibfield  {journal} {\bibinfo  {journal} {Nat. Electron.}\ }\textbf {\bibinfo {volume} {6}},\ \bibinfo {pages} {119} (\bibinfo {year} {2022})}\BibitemShut {NoStop}%
\bibitem [{\citenamefont {Lu}\ \emph {et~al.}(2024)\citenamefont {Lu}, \citenamefont {Lin}, \citenamefont {Pi}, \citenamefont {Zhang}, \citenamefont {Li}, \citenamefont {Gong}, \citenamefont {Yan}, \citenamefont {Ruan}, \citenamefont {Li}, \citenamefont {Zhang}, \citenamefont {Li}, \citenamefont {He}, \citenamefont {Wu}, \citenamefont {Zhang}, \citenamefont {Weng}, \citenamefont {Zeng},\ and\ \citenamefont {Xu}}]{RN1541}%
  \BibitemOpen
  \bibfield  {author} {\bibinfo {author} {\bibfnamefont {X.}~\bibnamefont {Lu}}, \bibinfo {author} {\bibfnamefont {Z.}~\bibnamefont {Lin}}, \bibinfo {author} {\bibfnamefont {H.}~\bibnamefont {Pi}}, \bibinfo {author} {\bibfnamefont {T.}~\bibnamefont {Zhang}}, \bibinfo {author} {\bibfnamefont {G.}~\bibnamefont {Li}}, \bibinfo {author} {\bibfnamefont {Y.}~\bibnamefont {Gong}}, \bibinfo {author} {\bibfnamefont {Y.}~\bibnamefont {Yan}}, \bibinfo {author} {\bibfnamefont {X.}~\bibnamefont {Ruan}}, \bibinfo {author} {\bibfnamefont {Y.}~\bibnamefont {Li}}, \bibinfo {author} {\bibfnamefont {H.}~\bibnamefont {Zhang}}, \bibinfo {author} {\bibfnamefont {L.}~\bibnamefont {Li}}, \bibinfo {author} {\bibfnamefont {L.}~\bibnamefont {He}}, \bibinfo {author} {\bibfnamefont {J.}~\bibnamefont {Wu}}, \bibinfo {author} {\bibfnamefont {R.}~\bibnamefont {Zhang}}, \bibinfo {author} {\bibfnamefont {H.}~\bibnamefont {Weng}}, \bibinfo {author} {\bibfnamefont {C.}~\bibnamefont {Zeng}}, \ and\ \bibinfo {author} {\bibfnamefont {Y.}~\bibnamefont {Xu}},\ }\bibfield  {title} {\enquote {\bibinfo {title} {Ultrafast magnetization enhancement via the dynamic spin-filter effect of type-ii weyl nodes in a kagome ferromagnet},}\ }\href@noop {} {\bibfield  {journal} {\bibinfo  {journal} {Nat. Commun.}\ }\textbf {\bibinfo {volume} {15}},\ \bibinfo {pages} {2410} (\bibinfo {year} {2024})}\BibitemShut {NoStop}%
\bibitem [{\citenamefont {Okamura}\ \emph {et~al.}(2020)\citenamefont {Okamura}, \citenamefont {Minami}, \citenamefont {Kato}, \citenamefont {Fujishiro}, \citenamefont {Kaneko}, \citenamefont {Ikeda}, \citenamefont {Muramoto}, \citenamefont {Kaneko}, \citenamefont {Ueda}, \citenamefont {Kocsis}, \citenamefont {Kanazawa}, \citenamefont {Taguchi}, \citenamefont {Koretsune}, \citenamefont {Fujiwara}, \citenamefont {Tsukazaki}, \citenamefont {Arita}, \citenamefont {Tokura},\ and\ \citenamefont {Takahashi}}]{RN1229}%
  \BibitemOpen
  \bibfield  {author} {\bibinfo {author} {\bibfnamefont {Y.}~\bibnamefont {Okamura}}, \bibinfo {author} {\bibfnamefont {S.}~\bibnamefont {Minami}}, \bibinfo {author} {\bibfnamefont {Y.}~\bibnamefont {Kato}}, \bibinfo {author} {\bibfnamefont {Y.}~\bibnamefont {Fujishiro}}, \bibinfo {author} {\bibfnamefont {Y.}~\bibnamefont {Kaneko}}, \bibinfo {author} {\bibfnamefont {J.}~\bibnamefont {Ikeda}}, \bibinfo {author} {\bibfnamefont {J.}~\bibnamefont {Muramoto}}, \bibinfo {author} {\bibfnamefont {R.}~\bibnamefont {Kaneko}}, \bibinfo {author} {\bibfnamefont {K.}~\bibnamefont {Ueda}}, \bibinfo {author} {\bibfnamefont {V.}~\bibnamefont {Kocsis}}, \bibinfo {author} {\bibfnamefont {N.}~\bibnamefont {Kanazawa}}, \bibinfo {author} {\bibfnamefont {Y.}~\bibnamefont {Taguchi}}, \bibinfo {author} {\bibfnamefont {T.}~\bibnamefont {Koretsune}}, \bibinfo {author} {\bibfnamefont {K.}~\bibnamefont {Fujiwara}}, \bibinfo {author} {\bibfnamefont {A.}~\bibnamefont {Tsukazaki}}, \bibinfo {author} {\bibfnamefont {R.}~\bibnamefont {Arita}}, \bibinfo {author} {\bibfnamefont {Y.}~\bibnamefont {Tokura}}, \ and\ \bibinfo {author} {\bibfnamefont {Y.}~\bibnamefont {Takahashi}},\ }\bibfield  {title} {\enquote {\bibinfo {title} {Giant magneto-optical responses in magnetic weyl semimetal co$_3$sn$_2$s$_2$},}\ }\href@noop {} {\bibfield  {journal} {\bibinfo  {journal} {Nat. Commun.}\ }\textbf {\bibinfo {volume} {11}},\ \bibinfo {pages} {4619} (\bibinfo {year} {2020})}\BibitemShut {NoStop}%
\bibitem [{\citenamefont {Li}\ \emph {et~al.}(2019)\citenamefont {Li}, \citenamefont {Xu}, \citenamefont {Shi}, \citenamefont {Fu}, \citenamefont {Jiao}, \citenamefont {Kamminga}, \citenamefont {Yu}, \citenamefont {Tüysüz}, \citenamefont {Kumar}, \citenamefont {Süß}, \citenamefont {Saha}, \citenamefont {Srivastava}, \citenamefont {Wirth}, \citenamefont {Auffermann}, \citenamefont {Gooth}, \citenamefont {Parkin}, \citenamefont {Sun}, \citenamefont {Liu},\ and\ \citenamefont {Felser}}]{RN598}%
  \BibitemOpen
  \bibfield  {author} {\bibinfo {author} {\bibfnamefont {G.}~\bibnamefont {Li}}, \bibinfo {author} {\bibfnamefont {Q.}~\bibnamefont {Xu}}, \bibinfo {author} {\bibfnamefont {W.}~\bibnamefont {Shi}}, \bibinfo {author} {\bibfnamefont {C.}~\bibnamefont {Fu}}, \bibinfo {author} {\bibfnamefont {L.}~\bibnamefont {Jiao}}, \bibinfo {author} {\bibfnamefont {M.~E.}\ \bibnamefont {Kamminga}}, \bibinfo {author} {\bibfnamefont {M.}~\bibnamefont {Yu}}, \bibinfo {author} {\bibfnamefont {H.}~\bibnamefont {Tüysüz}}, \bibinfo {author} {\bibfnamefont {N.}~\bibnamefont {Kumar}}, \bibinfo {author} {\bibfnamefont {V.}~\bibnamefont {Süß}}, \bibinfo {author} {\bibfnamefont {R.}~\bibnamefont {Saha}}, \bibinfo {author} {\bibfnamefont {A.~K.}\ \bibnamefont {Srivastava}}, \bibinfo {author} {\bibfnamefont {S.}~\bibnamefont {Wirth}}, \bibinfo {author} {\bibfnamefont {G.}~\bibnamefont {Auffermann}}, \bibinfo {author} {\bibfnamefont {J.}~\bibnamefont {Gooth}}, \bibinfo {author} {\bibfnamefont {S.}~\bibnamefont {Parkin}}, \bibinfo {author} {\bibfnamefont {Y.}~\bibnamefont {Sun}}, \bibinfo {author} {\bibfnamefont {E.}~\bibnamefont {Liu}}, \ and\ \bibinfo {author} {\bibfnamefont {C.}~\bibnamefont {Felser}},\ }\bibfield  {title} {\enquote {\bibinfo {title} {Surface states in bulk single crystal of topological semimetal co$_3$sn$_2$s$_2$ toward water oxidation},}\ }\href@noop {} {\bibfield  {journal} {\bibinfo  {journal} {Sci. Adv.}\ }\textbf {\bibinfo {volume} {5}},\ \bibinfo {pages} {eaaw9867} (\bibinfo {year} {2019})}\BibitemShut {NoStop}%
\bibitem [{\citenamefont {Zhao}\ \emph {et~al.}(2022)\citenamefont {Zhao}, \citenamefont {Zhu}, \citenamefont {Jiang}, \citenamefont {Li}, \citenamefont {Meng}, \citenamefont {Guo}, \citenamefont {Li}, \citenamefont {Huang}, \citenamefont {Zhang}, \citenamefont {Zhang}, \citenamefont {Ho}, \citenamefont {Zhang}, \citenamefont {Liu},\ and\ \citenamefont {Zhi}}]{RN1538}%
  \BibitemOpen
  \bibfield  {author} {\bibinfo {author} {\bibfnamefont {Y.}~\bibnamefont {Zhao}}, \bibinfo {author} {\bibfnamefont {Y.}~\bibnamefont {Zhu}}, \bibinfo {author} {\bibfnamefont {F.}~\bibnamefont {Jiang}}, \bibinfo {author} {\bibfnamefont {Y.}~\bibnamefont {Li}}, \bibinfo {author} {\bibfnamefont {Y.}~\bibnamefont {Meng}}, \bibinfo {author} {\bibfnamefont {Y.}~\bibnamefont {Guo}}, \bibinfo {author} {\bibfnamefont {Q.}~\bibnamefont {Li}}, \bibinfo {author} {\bibfnamefont {Z.}~\bibnamefont {Huang}}, \bibinfo {author} {\bibfnamefont {S.}~\bibnamefont {Zhang}}, \bibinfo {author} {\bibfnamefont {R.}~\bibnamefont {Zhang}}, \bibinfo {author} {\bibfnamefont {J.~C.}\ \bibnamefont {Ho}}, \bibinfo {author} {\bibfnamefont {Q.}~\bibnamefont {Zhang}}, \bibinfo {author} {\bibfnamefont {W.}~\bibnamefont {Liu}}, \ and\ \bibinfo {author} {\bibfnamefont {C.}~\bibnamefont {Zhi}},\ }\bibfield  {title} {\enquote {\bibinfo {title} {Vacancy modulating co3sn2s2 topological semimetal for aqueous zinc-ion batteries},}\ }\href@noop {} {\bibfield  {journal} {\bibinfo  {journal} {Angew. Chem. Int. Ed. Engl.}\ }\textbf {\bibinfo {volume} {61}},\ \bibinfo {pages} {e202111826} (\bibinfo {year} {2022})}\BibitemShut {NoStop}%
\bibitem [{\citenamefont {Ghimire}\ \emph {et~al.}(2019)\citenamefont {Ghimire}, \citenamefont {Facio}, \citenamefont {You}, \citenamefont {Ye}, \citenamefont {Checkelsky}, \citenamefont {Fang}, \citenamefont {Kaxiras}, \citenamefont {Richter},\ and\ \citenamefont {van~den Brink}}]{RN1411}%
  \BibitemOpen
  \bibfield  {author} {\bibinfo {author} {\bibfnamefont {M.~P.}\ \bibnamefont {Ghimire}}, \bibinfo {author} {\bibfnamefont {J.~I.}\ \bibnamefont {Facio}}, \bibinfo {author} {\bibfnamefont {J.-S.}\ \bibnamefont {You}}, \bibinfo {author} {\bibfnamefont {L.}~\bibnamefont {Ye}}, \bibinfo {author} {\bibfnamefont {J.~G.}\ \bibnamefont {Checkelsky}}, \bibinfo {author} {\bibfnamefont {S.}~\bibnamefont {Fang}}, \bibinfo {author} {\bibfnamefont {E.}~\bibnamefont {Kaxiras}}, \bibinfo {author} {\bibfnamefont {M.}~\bibnamefont {Richter}}, \ and\ \bibinfo {author} {\bibfnamefont {J.}~\bibnamefont {van~den Brink}},\ }\bibfield  {title} {\enquote {\bibinfo {title} {Creating weyl nodes and controlling their energy by magnetization rotation},}\ }\href@noop {} {\bibfield  {journal} {\bibinfo  {journal} {Phys. Rev. Research}\ }\textbf {\bibinfo {volume} {1}},\ \bibinfo {pages} {032044(R)} (\bibinfo {year} {2019})}\BibitemShut {NoStop}%
\bibitem [{\citenamefont {Nakazawa}, \citenamefont {Kato},\ and\ \citenamefont {Motome}(2024)}]{RN1547}%
  \BibitemOpen
  \bibfield  {author} {\bibinfo {author} {\bibfnamefont {K.}~\bibnamefont {Nakazawa}}, \bibinfo {author} {\bibfnamefont {Y.}~\bibnamefont {Kato}}, \ and\ \bibinfo {author} {\bibfnamefont {Y.}~\bibnamefont {Motome}},\ }\bibfield  {title} {\enquote {\bibinfo {title} {Topological transitions by magnetization rotation in kagome monolayers of the ferromagnetic weyl semimetal co-based shandite},}\ }\href@noop {} {\bibfield  {journal} {\bibinfo  {journal} {Phys. Rev. B}\ }\textbf {\bibinfo {volume} {110}},\ \bibinfo {pages} {085112} (\bibinfo {year} {2024})}\BibitemShut {NoStop}%
\bibitem [{\citenamefont {Yang}\ \emph {et~al.}(2020)\citenamefont {Yang}, \citenamefont {Zhang}, \citenamefont {Zhou}, \citenamefont {Dai}, \citenamefont {Liao}, \citenamefont {Weng},\ and\ \citenamefont {Qiu}}]{RN772}%
  \BibitemOpen
  \bibfield  {author} {\bibinfo {author} {\bibfnamefont {R.}~\bibnamefont {Yang}}, \bibinfo {author} {\bibfnamefont {T.}~\bibnamefont {Zhang}}, \bibinfo {author} {\bibfnamefont {L.}~\bibnamefont {Zhou}}, \bibinfo {author} {\bibfnamefont {Y.}~\bibnamefont {Dai}}, \bibinfo {author} {\bibfnamefont {Z.}~\bibnamefont {Liao}}, \bibinfo {author} {\bibfnamefont {H.}~\bibnamefont {Weng}}, \ and\ \bibinfo {author} {\bibfnamefont {X.}~\bibnamefont {Qiu}},\ }\bibfield  {title} {\enquote {\bibinfo {title} {Magnetization-induced band shift in ferromagnetic weyl semimetal co$_3$sn$_2$s$_2$},}\ }\href@noop {} {\bibfield  {journal} {\bibinfo  {journal} {Phys. Rev. Lett.}\ }\textbf {\bibinfo {volume} {124}},\ \bibinfo {pages} {077403} (\bibinfo {year} {2020})}\BibitemShut {NoStop}%
\bibitem [{\citenamefont {Zhang}\ \emph {et~al.}(2021)\citenamefont {Zhang}, \citenamefont {Okamoto}, \citenamefont {Samolyuk}, \citenamefont {Stone}, \citenamefont {Kolesnikov}, \citenamefont {Xue}, \citenamefont {Yan}, \citenamefont {McGuire}, \citenamefont {Mandrus},\ and\ \citenamefont {Tennant}}]{RN1537}%
  \BibitemOpen
  \bibfield  {author} {\bibinfo {author} {\bibfnamefont {Q.}~\bibnamefont {Zhang}}, \bibinfo {author} {\bibfnamefont {S.}~\bibnamefont {Okamoto}}, \bibinfo {author} {\bibfnamefont {G.~D.}\ \bibnamefont {Samolyuk}}, \bibinfo {author} {\bibfnamefont {M.~B.}\ \bibnamefont {Stone}}, \bibinfo {author} {\bibfnamefont {A.~I.}\ \bibnamefont {Kolesnikov}}, \bibinfo {author} {\bibfnamefont {R.}~\bibnamefont {Xue}}, \bibinfo {author} {\bibfnamefont {J.}~\bibnamefont {Yan}}, \bibinfo {author} {\bibfnamefont {M.~A.}\ \bibnamefont {McGuire}}, \bibinfo {author} {\bibfnamefont {D.}~\bibnamefont {Mandrus}}, \ and\ \bibinfo {author} {\bibfnamefont {D.~A.}\ \bibnamefont {Tennant}},\ }\bibfield  {title} {\enquote {\bibinfo {title} {Unusual exchange couplings and intermediate temperature weyl state in co$_3$sn$_2$s$_2$},}\ }\href@noop {} {\bibfield  {journal} {\bibinfo  {journal} {Phys. Rev. Lett.}\ }\textbf {\bibinfo {volume} {127}},\ \bibinfo {pages} {117201} (\bibinfo {year} {2021})}\BibitemShut {NoStop}%
\bibitem [{\citenamefont {Soh}\ \emph {et~al.}(2022)\citenamefont {Soh}, \citenamefont {Yi}, \citenamefont {Zivkovic}, \citenamefont {Qureshi}, \citenamefont {Stunault}, \citenamefont {Ouladdiaf}, \citenamefont {Rodríguez-Velamazán}, \citenamefont {Shi}, \citenamefont {Rønnow},\ and\ \citenamefont {Boothroyd}}]{RN1540}%
  \BibitemOpen
  \bibfield  {author} {\bibinfo {author} {\bibfnamefont {J.-R.}\ \bibnamefont {Soh}}, \bibinfo {author} {\bibfnamefont {C.}~\bibnamefont {Yi}}, \bibinfo {author} {\bibfnamefont {I.}~\bibnamefont {Zivkovic}}, \bibinfo {author} {\bibfnamefont {N.}~\bibnamefont {Qureshi}}, \bibinfo {author} {\bibfnamefont {A.}~\bibnamefont {Stunault}}, \bibinfo {author} {\bibfnamefont {B.}~\bibnamefont {Ouladdiaf}}, \bibinfo {author} {\bibfnamefont {J.~A.}\ \bibnamefont {Rodríguez-Velamazán}}, \bibinfo {author} {\bibfnamefont {Y.}~\bibnamefont {Shi}}, \bibinfo {author} {\bibfnamefont {H.~M.}\ \bibnamefont {Rønnow}}, \ and\ \bibinfo {author} {\bibfnamefont {A.~T.}\ \bibnamefont {Boothroyd}},\ }\bibfield  {title} {\enquote {\bibinfo {title} {Magnetic structure of the topological semimetal co$_3$sn$_2$s$_2$},}\ }\href@noop {} {\bibfield  {journal} {\bibinfo  {journal} {Phys. Rev. B}\ }\textbf {\bibinfo {volume} {105}},\ \bibinfo {pages} {094435} (\bibinfo {year} {2022})}\BibitemShut {NoStop}%
\bibitem [{\citenamefont {Živković}\ \emph {et~al.}(2022)\citenamefont {Živković}, \citenamefont {Yadav}, \citenamefont {Soh}, \citenamefont {Yi}, \citenamefont {Shi}, \citenamefont {Yazyev},\ and\ \citenamefont {Rønnow}}]{RN1544}%
  \BibitemOpen
  \bibfield  {author} {\bibinfo {author} {\bibfnamefont {I.}~\bibnamefont {Živković}}, \bibinfo {author} {\bibfnamefont {R.}~\bibnamefont {Yadav}}, \bibinfo {author} {\bibfnamefont {J.-R.}\ \bibnamefont {Soh}}, \bibinfo {author} {\bibfnamefont {C.}~\bibnamefont {Yi}}, \bibinfo {author} {\bibfnamefont {Y.}~\bibnamefont {Shi}}, \bibinfo {author} {\bibfnamefont {O.~V.}\ \bibnamefont {Yazyev}}, \ and\ \bibinfo {author} {\bibfnamefont {H.~M.}\ \bibnamefont {Rønnow}},\ }\bibfield  {title} {\enquote {\bibinfo {title} {Unraveling the origin of the peculiar transition in the magnetically ordered phase of the weyl semimetal co$_3$sn$_2$s$_2$},}\ }\href@noop {} {\bibfield  {journal} {\bibinfo  {journal} {Phys. Rev. B}\ }\textbf {\bibinfo {volume} {106}},\ \bibinfo {pages} {L180403} (\bibinfo {year} {2022})}\BibitemShut {NoStop}%
\bibitem [{\citenamefont {Howlader}\ \emph {et~al.}(2020)\citenamefont {Howlader}, \citenamefont {Ramachandran}, \citenamefont {Monga}, \citenamefont {Singh},\ and\ \citenamefont {Sheet}}]{RN1418}%
  \BibitemOpen
  \bibfield  {author} {\bibinfo {author} {\bibfnamefont {S.}~\bibnamefont {Howlader}}, \bibinfo {author} {\bibfnamefont {R.}~\bibnamefont {Ramachandran}}, \bibinfo {author} {\bibfnamefont {S.}~\bibnamefont {Monga}}, \bibinfo {author} {\bibfnamefont {Y.}~\bibnamefont {Singh}}, \ and\ \bibinfo {author} {\bibfnamefont {G.}~\bibnamefont {Sheet}},\ }\bibfield  {title} {\enquote {\bibinfo {title} {Domain structure evolution in the ferromagnetic kagome-lattice weyl semimetal co$_3$sn$_2$s$_2$},}\ }\href@noop {} {\bibfield  {journal} {\bibinfo  {journal} {J Phys Condens Matter}\ }\textbf {\bibinfo {volume} {33}},\ \bibinfo {pages} {075801} (\bibinfo {year} {2020})}\BibitemShut {NoStop}%
\bibitem [{\citenamefont {Lee}\ \emph {et~al.}(2022)\citenamefont {Lee}, \citenamefont {Vir}, \citenamefont {Manna}, \citenamefont {Shekhar}, \citenamefont {Moore}, \citenamefont {Kastner}, \citenamefont {Felser},\ and\ \citenamefont {Orenstein}}]{RN1502}%
  \BibitemOpen
  \bibfield  {author} {\bibinfo {author} {\bibfnamefont {C.}~\bibnamefont {Lee}}, \bibinfo {author} {\bibfnamefont {P.}~\bibnamefont {Vir}}, \bibinfo {author} {\bibfnamefont {K.}~\bibnamefont {Manna}}, \bibinfo {author} {\bibfnamefont {C.}~\bibnamefont {Shekhar}}, \bibinfo {author} {\bibfnamefont {J.~E.}\ \bibnamefont {Moore}}, \bibinfo {author} {\bibfnamefont {M.~A.}\ \bibnamefont {Kastner}}, \bibinfo {author} {\bibfnamefont {C.}~\bibnamefont {Felser}}, \ and\ \bibinfo {author} {\bibfnamefont {J.}~\bibnamefont {Orenstein}},\ }\bibfield  {title} {\enquote {\bibinfo {title} {Observation of a phase transition within the domain walls of ferromagnetic co$_3$sn$_2$s$_2$},}\ }\href@noop {} {\bibfield  {journal} {\bibinfo  {journal} {Nat. Commun.}\ }\textbf {\bibinfo {volume} {13}},\ \bibinfo {pages} {3000} (\bibinfo {year} {2022})}\BibitemShut {NoStop}%
\bibitem [{\citenamefont {Sugawara}\ \emph {et~al.}(2019)\citenamefont {Sugawara}, \citenamefont {Akashi}, \citenamefont {Kassem}, \citenamefont {Tabata}, \citenamefont {Waki},\ and\ \citenamefont {Nakamura}}]{RN594}%
  \BibitemOpen
  \bibfield  {author} {\bibinfo {author} {\bibfnamefont {A.}~\bibnamefont {Sugawara}}, \bibinfo {author} {\bibfnamefont {T.}~\bibnamefont {Akashi}}, \bibinfo {author} {\bibfnamefont {M.~A.}\ \bibnamefont {Kassem}}, \bibinfo {author} {\bibfnamefont {Y.}~\bibnamefont {Tabata}}, \bibinfo {author} {\bibfnamefont {T.}~\bibnamefont {Waki}}, \ and\ \bibinfo {author} {\bibfnamefont {H.}~\bibnamefont {Nakamura}},\ }\bibfield  {title} {\enquote {\bibinfo {title} {Magnetic domain structure within half-metallic ferromagnetic kagome compound co$_3$sn$_2$s$_2$},}\ }\href@noop {} {\bibfield  {journal} {\bibinfo  {journal} {Phys. Rev. Mater.}\ }\textbf {\bibinfo {volume} {3}},\ \bibinfo {pages} {104421} (\bibinfo {year} {2019})}\BibitemShut {NoStop}%
\bibitem [{\citenamefont {Chen}\ \emph {et~al.}(2019)\citenamefont {Chen}, \citenamefont {Wang}, \citenamefont {Gu}, \citenamefont {Wang}, \citenamefont {Zhou}, \citenamefont {An}, \citenamefont {Zhou}, \citenamefont {Zhang}, \citenamefont {Chen}, \citenamefont {Yuan}, \citenamefont {Qi}, \citenamefont {Zhang}, \citenamefont {Zhou}, \citenamefont {Zhou}, \citenamefont {Yao},\ and\ \citenamefont {Yang}}]{RN552}%
  \BibitemOpen
  \bibfield  {author} {\bibinfo {author} {\bibfnamefont {X.}~\bibnamefont {Chen}}, \bibinfo {author} {\bibfnamefont {M.}~\bibnamefont {Wang}}, \bibinfo {author} {\bibfnamefont {C.}~\bibnamefont {Gu}}, \bibinfo {author} {\bibfnamefont {S.}~\bibnamefont {Wang}}, \bibinfo {author} {\bibfnamefont {Y.}~\bibnamefont {Zhou}}, \bibinfo {author} {\bibfnamefont {C.}~\bibnamefont {An}}, \bibinfo {author} {\bibfnamefont {Y.}~\bibnamefont {Zhou}}, \bibinfo {author} {\bibfnamefont {B.}~\bibnamefont {Zhang}}, \bibinfo {author} {\bibfnamefont {C.}~\bibnamefont {Chen}}, \bibinfo {author} {\bibfnamefont {Y.}~\bibnamefont {Yuan}}, \bibinfo {author} {\bibfnamefont {M.}~\bibnamefont {Qi}}, \bibinfo {author} {\bibfnamefont {L.}~\bibnamefont {Zhang}}, \bibinfo {author} {\bibfnamefont {H.}~\bibnamefont {Zhou}}, \bibinfo {author} {\bibfnamefont {J.}~\bibnamefont {Zhou}}, \bibinfo {author} {\bibfnamefont {Y.}~\bibnamefont {Yao}}, \ and\ \bibinfo {author} {\bibfnamefont {Z.}~\bibnamefont {Yang}},\ }\bibfield  {title} {\enquote {\bibinfo {title} {Pressure-tunable large anomalous hall effect of the ferromagnetic kagome-lattice weyl semimetal co$_3$sn$_2$s$_2$},}\ }\href@noop {} {\bibfield  {journal} {\bibinfo  {journal} {Phys. Rev. B}\ }\textbf {\bibinfo {volume} {100}},\ \bibinfo {pages} {165145} (\bibinfo {year} {2019})}\BibitemShut {NoStop}%
\bibitem [{\citenamefont {Guguchia}\ \emph {et~al.}(2020)\citenamefont {Guguchia}, \citenamefont {Verezhak}, \citenamefont {Gawryluk}, \citenamefont {Tsirkin}, \citenamefont {Yin}, \citenamefont {Belopolski}, \citenamefont {Zhou}, \citenamefont {Simutis}, \citenamefont {Zhang}, \citenamefont {Cochran}, \citenamefont {Chang}, \citenamefont {Pomjakushina}, \citenamefont {Keller}, \citenamefont {Skrzeczkowska}, \citenamefont {Wang}, \citenamefont {Lei}, \citenamefont {Khasanov}, \citenamefont {Amato}, \citenamefont {Jia}, \citenamefont {Neupert}, \citenamefont {Luetkens},\ and\ \citenamefont {Hasan}}]{RN764}%
  \BibitemOpen
  \bibfield  {author} {\bibinfo {author} {\bibfnamefont {Z.}~\bibnamefont {Guguchia}}, \bibinfo {author} {\bibfnamefont {J.~A.~T.}\ \bibnamefont {Verezhak}}, \bibinfo {author} {\bibfnamefont {D.~J.}\ \bibnamefont {Gawryluk}}, \bibinfo {author} {\bibfnamefont {S.~S.}\ \bibnamefont {Tsirkin}}, \bibinfo {author} {\bibfnamefont {J.~X.}\ \bibnamefont {Yin}}, \bibinfo {author} {\bibfnamefont {I.}~\bibnamefont {Belopolski}}, \bibinfo {author} {\bibfnamefont {H.}~\bibnamefont {Zhou}}, \bibinfo {author} {\bibfnamefont {G.}~\bibnamefont {Simutis}}, \bibinfo {author} {\bibfnamefont {S.~S.}\ \bibnamefont {Zhang}}, \bibinfo {author} {\bibfnamefont {T.~A.}\ \bibnamefont {Cochran}}, \bibinfo {author} {\bibfnamefont {G.}~\bibnamefont {Chang}}, \bibinfo {author} {\bibfnamefont {E.}~\bibnamefont {Pomjakushina}}, \bibinfo {author} {\bibfnamefont {L.}~\bibnamefont {Keller}}, \bibinfo {author} {\bibfnamefont {Z.}~\bibnamefont {Skrzeczkowska}}, \bibinfo {author} {\bibfnamefont {Q.}~\bibnamefont {Wang}}, \bibinfo {author} {\bibfnamefont {H.~C.}\ \bibnamefont {Lei}}, \bibinfo {author} {\bibfnamefont {R.}~\bibnamefont {Khasanov}}, \bibinfo {author} {\bibfnamefont {A.}~\bibnamefont {Amato}}, \bibinfo {author} {\bibfnamefont {S.}~\bibnamefont {Jia}}, \bibinfo {author} {\bibfnamefont {T.}~\bibnamefont {Neupert}}, \bibinfo {author} {\bibfnamefont {H.}~\bibnamefont {Luetkens}}, \ and\ \bibinfo {author} {\bibfnamefont {M.~Z.}\ \bibnamefont {Hasan}},\ }\bibfield  {title} {\enquote {\bibinfo {title} {Tunable anomalous hall conductivity through volume-wise magnetic competition in a topological kagome magnet},}\ }\href@noop {} {\bibfield  {journal} {\bibinfo  {journal} {Nat. Commun.}\ }\textbf {\bibinfo {volume} {11}},\ \bibinfo {pages} {559} (\bibinfo {year} {2020})}\BibitemShut {NoStop}%
\bibitem [{\citenamefont {Pate}\ \emph {et~al.}(2023)\citenamefont {Pate}, \citenamefont {Wang}, \citenamefont {Shen}, \citenamefont {Jiang}, \citenamefont {Welp}, \citenamefont {Kwok}, \citenamefont {Xu}, \citenamefont {Li}, \citenamefont {Divan},\ and\ \citenamefont {Xiao}}]{RN1546}%
  \BibitemOpen
  \bibfield  {author} {\bibinfo {author} {\bibfnamefont {S.~E.}\ \bibnamefont {Pate}}, \bibinfo {author} {\bibfnamefont {B.}~\bibnamefont {Wang}}, \bibinfo {author} {\bibfnamefont {B.}~\bibnamefont {Shen}}, \bibinfo {author} {\bibfnamefont {J.~S.}\ \bibnamefont {Jiang}}, \bibinfo {author} {\bibfnamefont {U.}~\bibnamefont {Welp}}, \bibinfo {author} {\bibfnamefont {W.-K.}\ \bibnamefont {Kwok}}, \bibinfo {author} {\bibfnamefont {J.}~\bibnamefont {Xu}}, \bibinfo {author} {\bibfnamefont {K.}~\bibnamefont {Li}}, \bibinfo {author} {\bibfnamefont {R.}~\bibnamefont {Divan}}, \ and\ \bibinfo {author} {\bibfnamefont {Z.-L.}\ \bibnamefont {Xiao}},\ }\bibfield  {title} {\enquote {\bibinfo {title} {Field orientation dependent magnetic phases in the weyl semimetal co$_3$sn$_2$s$_2$},}\ }\href@noop {} {\bibfield  {journal} {\bibinfo  {journal} {Phys. Rev. B}\ }\textbf {\bibinfo {volume} {108}},\ \bibinfo {pages} {L100408} (\bibinfo {year} {2023})}\BibitemShut {NoStop}%
\bibitem [{\citenamefont {Wu}\ \emph {et~al.}(2020)\citenamefont {Wu}, \citenamefont {Sun}, \citenamefont {Hsieh}, \citenamefont {Chen}, \citenamefont {Kakarla}, \citenamefont {Deng}, \citenamefont {Chu},\ and\ \citenamefont {Yang}}]{RN1417}%
  \BibitemOpen
  \bibfield  {author} {\bibinfo {author} {\bibfnamefont {H.~C.}\ \bibnamefont {Wu}}, \bibinfo {author} {\bibfnamefont {P.~J.}\ \bibnamefont {Sun}}, \bibinfo {author} {\bibfnamefont {D.~J.}\ \bibnamefont {Hsieh}}, \bibinfo {author} {\bibfnamefont {H.~J.}\ \bibnamefont {Chen}}, \bibinfo {author} {\bibfnamefont {D.~C.}\ \bibnamefont {Kakarla}}, \bibinfo {author} {\bibfnamefont {L.~Z.}\ \bibnamefont {Deng}}, \bibinfo {author} {\bibfnamefont {C.~W.}\ \bibnamefont {Chu}}, \ and\ \bibinfo {author} {\bibfnamefont {H.~D.}\ \bibnamefont {Yang}},\ }\bibfield  {title} {\enquote {\bibinfo {title} {Observation of skyrmion-like magnetism in magnetic weyl semimetal co$_3$sn$_2$s$_2$},}\ }\href@noop {} {\bibfield  {journal} {\bibinfo  {journal} {Mater. Today Phys.}\ }\textbf {\bibinfo {volume} {12}},\ \bibinfo {pages} {100189} (\bibinfo {year} {2020})}\BibitemShut {NoStop}%
\bibitem [{\citenamefont {Lachman}\ \emph {et~al.}(2020)\citenamefont {Lachman}, \citenamefont {Murphy}, \citenamefont {Maksimovic}, \citenamefont {Kealhofer}, \citenamefont {Haley}, \citenamefont {McDonald}, \citenamefont {Long},\ and\ \citenamefont {Analytis}}]{RN884}%
  \BibitemOpen
  \bibfield  {author} {\bibinfo {author} {\bibfnamefont {E.}~\bibnamefont {Lachman}}, \bibinfo {author} {\bibfnamefont {R.~A.}\ \bibnamefont {Murphy}}, \bibinfo {author} {\bibfnamefont {N.}~\bibnamefont {Maksimovic}}, \bibinfo {author} {\bibfnamefont {R.}~\bibnamefont {Kealhofer}}, \bibinfo {author} {\bibfnamefont {S.}~\bibnamefont {Haley}}, \bibinfo {author} {\bibfnamefont {R.~D.}\ \bibnamefont {McDonald}}, \bibinfo {author} {\bibfnamefont {J.~R.}\ \bibnamefont {Long}}, \ and\ \bibinfo {author} {\bibfnamefont {J.~G.}\ \bibnamefont {Analytis}},\ }\bibfield  {title} {\enquote {\bibinfo {title} {Exchange biased anomalous hall effect driven by frustration in a magnetic kagome lattice},}\ }\href@noop {} {\bibfield  {journal} {\bibinfo  {journal} {Nat. Commun.}\ }\textbf {\bibinfo {volume} {11}},\ \bibinfo {pages} {560} (\bibinfo {year} {2020})}\BibitemShut {NoStop}%
\bibitem [{\citenamefont {Noah}\ \emph {et~al.}(2022)\citenamefont {Noah}, \citenamefont {Toric}, \citenamefont {Feld}, \citenamefont {Zissman}, \citenamefont {Gutfreund}, \citenamefont {Tsruya}, \citenamefont {Devidas}, \citenamefont {Alpern}, \citenamefont {Vakahi}, \citenamefont {Steinberg}, \citenamefont {Huber}, \citenamefont {Analytis}, \citenamefont {Gazit}, \citenamefont {Lachman},\ and\ \citenamefont {Anahory}}]{RN1486}%
  \BibitemOpen
  \bibfield  {author} {\bibinfo {author} {\bibfnamefont {A.}~\bibnamefont {Noah}}, \bibinfo {author} {\bibfnamefont {F.}~\bibnamefont {Toric}}, \bibinfo {author} {\bibfnamefont {T.~D.}\ \bibnamefont {Feld}}, \bibinfo {author} {\bibfnamefont {G.}~\bibnamefont {Zissman}}, \bibinfo {author} {\bibfnamefont {A.}~\bibnamefont {Gutfreund}}, \bibinfo {author} {\bibfnamefont {D.}~\bibnamefont {Tsruya}}, \bibinfo {author} {\bibfnamefont {T.~R.}\ \bibnamefont {Devidas}}, \bibinfo {author} {\bibfnamefont {H.}~\bibnamefont {Alpern}}, \bibinfo {author} {\bibfnamefont {A.}~\bibnamefont {Vakahi}}, \bibinfo {author} {\bibfnamefont {H.}~\bibnamefont {Steinberg}}, \bibinfo {author} {\bibfnamefont {M.~E.}\ \bibnamefont {Huber}}, \bibinfo {author} {\bibfnamefont {J.~G.}\ \bibnamefont {Analytis}}, \bibinfo {author} {\bibfnamefont {S.}~\bibnamefont {Gazit}}, \bibinfo {author} {\bibfnamefont {E.}~\bibnamefont {Lachman}}, \ and\ \bibinfo {author} {\bibfnamefont {Y.}~\bibnamefont {Anahory}},\ }\bibfield  {title} {\enquote {\bibinfo {title} {Tunable exchange bias in the magnetic weyl semimetal co$_3$sn$_2$s$_2$},}\ }\href@noop {} {\bibfield  {journal} {\bibinfo  {journal} {Phys. Rev. B}\ }\textbf {\bibinfo {volume} {105}},\ \bibinfo {pages} {144423} (\bibinfo {year} {2022})}\BibitemShut {NoStop}%
\bibitem [{\citenamefont {Guin}\ \emph {et~al.}(2019{\natexlab{b}})\citenamefont {Guin}, \citenamefont {Vir}, \citenamefont {Zhang}, \citenamefont {Kumar}, \citenamefont {Watzman}, \citenamefont {Fu}, \citenamefont {Liu}, \citenamefont {Manna}, \citenamefont {Schnelle}, \citenamefont {Gooth}, \citenamefont {Shekhar}, \citenamefont {Sun},\ and\ \citenamefont {Felser}}]{RN824}%
  \BibitemOpen
  \bibfield  {author} {\bibinfo {author} {\bibfnamefont {S.~N.}\ \bibnamefont {Guin}}, \bibinfo {author} {\bibfnamefont {P.}~\bibnamefont {Vir}}, \bibinfo {author} {\bibfnamefont {Y.}~\bibnamefont {Zhang}}, \bibinfo {author} {\bibfnamefont {N.}~\bibnamefont {Kumar}}, \bibinfo {author} {\bibfnamefont {S.~J.}\ \bibnamefont {Watzman}}, \bibinfo {author} {\bibfnamefont {C.}~\bibnamefont {Fu}}, \bibinfo {author} {\bibfnamefont {E.}~\bibnamefont {Liu}}, \bibinfo {author} {\bibfnamefont {K.}~\bibnamefont {Manna}}, \bibinfo {author} {\bibfnamefont {W.}~\bibnamefont {Schnelle}}, \bibinfo {author} {\bibfnamefont {J.}~\bibnamefont {Gooth}}, \bibinfo {author} {\bibfnamefont {C.}~\bibnamefont {Shekhar}}, \bibinfo {author} {\bibfnamefont {Y.}~\bibnamefont {Sun}}, \ and\ \bibinfo {author} {\bibfnamefont {C.}~\bibnamefont {Felser}},\ }\bibfield  {title} {\enquote {\bibinfo {title} {Zero-field nernst effect in a ferromagnetic kagome-lattice weyl-semimetal co$_3$sn$_2$s$_2$},}\ }\href@noop {} {\bibfield  {journal} {\bibinfo  {journal} {Adv. Mater.}\ }\textbf {\bibinfo {volume} {31}},\ \bibinfo {pages} {1806622} (\bibinfo {year} {2019}{\natexlab{b}})}\BibitemShut {NoStop}%
\bibitem [{RN1(2021)}]{RN1520}%
  \BibitemOpen
  \href@noop {} {\emph {\bibinfo {title} {Handbook of Magnetism and Magnetic Materials}}}\ (\bibinfo  {publisher} {Springer Cham},\ \bibinfo {year} {2021})\BibitemShut {NoStop}%
\bibitem [{\citenamefont {Duerrschnabel}\ \emph {et~al.}(2017)\citenamefont {Duerrschnabel}, \citenamefont {Yi}, \citenamefont {Uestuener}, \citenamefont {Liesegang}, \citenamefont {Katter}, \citenamefont {Kleebe}, \citenamefont {Xu}, \citenamefont {Gutfleisch},\ and\ \citenamefont {Molina-Luna}}]{RN1521}%
  \BibitemOpen
  \bibfield  {author} {\bibinfo {author} {\bibfnamefont {M.}~\bibnamefont {Duerrschnabel}}, \bibinfo {author} {\bibfnamefont {M.}~\bibnamefont {Yi}}, \bibinfo {author} {\bibfnamefont {K.}~\bibnamefont {Uestuener}}, \bibinfo {author} {\bibfnamefont {M.}~\bibnamefont {Liesegang}}, \bibinfo {author} {\bibfnamefont {M.}~\bibnamefont {Katter}}, \bibinfo {author} {\bibfnamefont {H.~J.}\ \bibnamefont {Kleebe}}, \bibinfo {author} {\bibfnamefont {B.}~\bibnamefont {Xu}}, \bibinfo {author} {\bibfnamefont {O.}~\bibnamefont {Gutfleisch}}, \ and\ \bibinfo {author} {\bibfnamefont {L.}~\bibnamefont {Molina-Luna}},\ }\bibfield  {title} {\enquote {\bibinfo {title} {Atomic structure and domain wall pinning in samarium-cobalt-based permanent magnets},}\ }\href@noop {} {\bibfield  {journal} {\bibinfo  {journal} {Nat. Commun.}\ }\textbf {\bibinfo {volume} {8}},\ \bibinfo {pages} {54} (\bibinfo {year} {2017})}\BibitemShut {NoStop}%
\bibitem [{\citenamefont {Gutfleisch}\ \emph {et~al.}(2006)\citenamefont {Gutfleisch}, \citenamefont {Müller}, \citenamefont {Khlopkov}, \citenamefont {Wolf}, \citenamefont {Yan}, \citenamefont {Schäfer}, \citenamefont {Gemming},\ and\ \citenamefont {Schultz}}]{RN1527}%
  \BibitemOpen
  \bibfield  {author} {\bibinfo {author} {\bibfnamefont {O.}~\bibnamefont {Gutfleisch}}, \bibinfo {author} {\bibfnamefont {K.~H.}\ \bibnamefont {Müller}}, \bibinfo {author} {\bibfnamefont {K.}~\bibnamefont {Khlopkov}}, \bibinfo {author} {\bibfnamefont {M.}~\bibnamefont {Wolf}}, \bibinfo {author} {\bibfnamefont {A.}~\bibnamefont {Yan}}, \bibinfo {author} {\bibfnamefont {R.}~\bibnamefont {Schäfer}}, \bibinfo {author} {\bibfnamefont {T.}~\bibnamefont {Gemming}}, \ and\ \bibinfo {author} {\bibfnamefont {L.}~\bibnamefont {Schultz}},\ }\bibfield  {title} {\enquote {\bibinfo {title} {Evolution of magnetic domain structures and coercivity in high-performance smco 2:17-type permanent magnets},}\ }\href@noop {} {\bibfield  {journal} {\bibinfo  {journal} {Acta Mater.}\ }\textbf {\bibinfo {volume} {54}},\ \bibinfo {pages} {997} (\bibinfo {year} {2006})}\BibitemShut {NoStop}%
\bibitem [{\citenamefont {Karpenkov}\ \emph {et~al.}(2022)\citenamefont {Karpenkov}, \citenamefont {Skokov}, \citenamefont {Dunaeva}, \citenamefont {Semenova}, \citenamefont {Lyakhova},\ and\ \citenamefont {Pastushenkov}}]{RN1530}%
  \BibitemOpen
  \bibfield  {author} {\bibinfo {author} {\bibfnamefont {A.~Y.}\ \bibnamefont {Karpenkov}}, \bibinfo {author} {\bibfnamefont {K.~P.}\ \bibnamefont {Skokov}}, \bibinfo {author} {\bibfnamefont {G.~G.}\ \bibnamefont {Dunaeva}}, \bibinfo {author} {\bibfnamefont {E.~M.}\ \bibnamefont {Semenova}}, \bibinfo {author} {\bibfnamefont {M.~B.}\ \bibnamefont {Lyakhova}}, \ and\ \bibinfo {author} {\bibfnamefont {Y.~G.}\ \bibnamefont {Pastushenkov}},\ }\bibfield  {title} {\enquote {\bibinfo {title} {Quantitative analyses of surface and bulk magnetization in nd$_2$fe$_14$b and smco$_5$ single crystals: towards understanding the large neff in nucleation-type magnets},}\ }\href@noop {} {\bibfield  {journal} {\bibinfo  {journal} {J. Phys. D: Appl. Phys.}\ }\textbf {\bibinfo {volume} {55}},\ \bibinfo {pages} {455002} (\bibinfo {year} {2022})}\BibitemShut {NoStop}%
\bibitem [{\citenamefont {Becker}(1973)}]{RN1518}%
  \BibitemOpen
  \bibfield  {author} {\bibinfo {author} {\bibfnamefont {J.}~\bibnamefont {Becker}},\ }\bibfield  {title} {\enquote {\bibinfo {title} {A model for the field dependence of magnetization discontinuities in high-anisotropy materials},}\ }\href@noop {} {\bibfield  {journal} {\bibinfo  {journal} {IEEE Trans. Magn.}\ }\textbf {\bibinfo {volume} {9}},\ \bibinfo {pages} {161} (\bibinfo {year} {1973})}\BibitemShut {NoStop}%
\bibitem [{\citenamefont {Zhang}\ \emph {et~al.}(2024)\citenamefont {Zhang}, \citenamefont {Aubert}, \citenamefont {Maccari}, \citenamefont {Dietz}, \citenamefont {Yue}, \citenamefont {Gutfleisch},\ and\ \citenamefont {Skokov}}]{RN1524}%
  \BibitemOpen
  \bibfield  {author} {\bibinfo {author} {\bibfnamefont {H.}~\bibnamefont {Zhang}}, \bibinfo {author} {\bibfnamefont {A.}~\bibnamefont {Aubert}}, \bibinfo {author} {\bibfnamefont {F.}~\bibnamefont {Maccari}}, \bibinfo {author} {\bibfnamefont {C.}~\bibnamefont {Dietz}}, \bibinfo {author} {\bibfnamefont {M.}~\bibnamefont {Yue}}, \bibinfo {author} {\bibfnamefont {O.}~\bibnamefont {Gutfleisch}}, \ and\ \bibinfo {author} {\bibfnamefont {K.}~\bibnamefont {Skokov}},\ }\bibfield  {title} {\enquote {\bibinfo {title} {Study of magnetization reversal and magnetic hardening in smco5 single crystal magnets},}\ }\href@noop {} {\bibfield  {journal} {\bibinfo  {journal} {J. Alloys Compd.}\ }\textbf {\bibinfo {volume} {993}},\ \bibinfo {pages} {174570} (\bibinfo {year} {2024})}\BibitemShut {NoStop}%
\bibitem [{\citenamefont {Chikazumi}(1986)}]{RN1529}%
  \BibitemOpen
  \bibfield  {author} {\bibinfo {author} {\bibfnamefont {S.}~\bibnamefont {Chikazumi}},\ }\bibfield  {title} {\enquote {\bibinfo {title} {Mechanism of high coercivity in rare-earth permanent magnets},}\ }\href@noop {} {\bibfield  {journal} {\bibinfo  {journal} {J. Magn. Magn. Mater.}\ }\textbf {\bibinfo {volume} {54-57}},\ \bibinfo {pages} {1551} (\bibinfo {year} {1986})}\BibitemShut {NoStop}%
\bibitem [{\citenamefont {Ener}\ \emph {et~al.}(2021)\citenamefont {Ener}, \citenamefont {Skokov}, \citenamefont {Palanisamy}, \citenamefont {Devillers}, \citenamefont {Fischbacher}, \citenamefont {Eslava}, \citenamefont {Maccari}, \citenamefont {Schäfer}, \citenamefont {Diop}, \citenamefont {Radulov}, \citenamefont {Gault}, \citenamefont {Hrkac}, \citenamefont {Dempsey}, \citenamefont {Schrefl}, \citenamefont {Raabe},\ and\ \citenamefont {Gutfleisch}}]{RN1551}%
  \BibitemOpen
  \bibfield  {author} {\bibinfo {author} {\bibfnamefont {S.}~\bibnamefont {Ener}}, \bibinfo {author} {\bibfnamefont {K.~P.}\ \bibnamefont {Skokov}}, \bibinfo {author} {\bibfnamefont {D.}~\bibnamefont {Palanisamy}}, \bibinfo {author} {\bibfnamefont {T.}~\bibnamefont {Devillers}}, \bibinfo {author} {\bibfnamefont {J.}~\bibnamefont {Fischbacher}}, \bibinfo {author} {\bibfnamefont {G.~G.}\ \bibnamefont {Eslava}}, \bibinfo {author} {\bibfnamefont {F.}~\bibnamefont {Maccari}}, \bibinfo {author} {\bibfnamefont {L.}~\bibnamefont {Schäfer}}, \bibinfo {author} {\bibfnamefont {L.~V.~B.}\ \bibnamefont {Diop}}, \bibinfo {author} {\bibfnamefont {I.}~\bibnamefont {Radulov}}, \bibinfo {author} {\bibfnamefont {B.}~\bibnamefont {Gault}}, \bibinfo {author} {\bibfnamefont {G.}~\bibnamefont {Hrkac}}, \bibinfo {author} {\bibfnamefont {N.~M.}\ \bibnamefont {Dempsey}}, \bibinfo {author} {\bibfnamefont {T.}~\bibnamefont {Schrefl}}, \bibinfo {author} {\bibfnamefont {D.}~\bibnamefont {Raabe}}, \ and\ \bibinfo {author} {\bibfnamefont {O.}~\bibnamefont {Gutfleisch}},\ }\bibfield  {title} {\enquote {\bibinfo {title} {Twins – a weak link in the magnetic hardening of thmn$_12$-type permanent magnets},}\ }\href@noop {} {\bibfield  {journal} {\bibinfo  {journal} {Acta Mater.}\ }\textbf {\bibinfo {volume} {214}},\ \bibinfo {pages} {116968} (\bibinfo {year} {2021})}\BibitemShut {NoStop}%
\bibitem [{\citenamefont {Katayama}\ \emph {et~al.}(1976)\citenamefont {Katayama}, \citenamefont {Ohkoshi}, \citenamefont {Koizumi}, \citenamefont {Shibata},\ and\ \citenamefont {Tsushima}}]{RN1523}%
  \BibitemOpen
  \bibfield  {author} {\bibinfo {author} {\bibfnamefont {T.}~\bibnamefont {Katayama}}, \bibinfo {author} {\bibfnamefont {M.}~\bibnamefont {Ohkoshi}}, \bibinfo {author} {\bibfnamefont {Y.}~\bibnamefont {Koizumi}}, \bibinfo {author} {\bibfnamefont {T.}~\bibnamefont {Shibata}}, \ and\ \bibinfo {author} {\bibfnamefont {T.}~\bibnamefont {Tsushima}},\ }\bibfield  {title} {\enquote {\bibinfo {title} {Magnetization reversal in gdco$_5$ single crystals},}\ }\href@noop {} {\bibfield  {journal} {\bibinfo  {journal} {Appl. Phys. Lett.}\ }\textbf {\bibinfo {volume} {28}},\ \bibinfo {pages} {635} (\bibinfo {year} {1976})}\BibitemShut {NoStop}%
\bibitem [{\citenamefont {Li}\ \emph {et~al.}(2022)\citenamefont {Li}, \citenamefont {Zhang}, \citenamefont {Fu}, \citenamefont {Wang}, \citenamefont {Wei},\ and\ \citenamefont {Bai}}]{RN1550}%
  \BibitemOpen
  \bibfield  {author} {\bibinfo {author} {\bibfnamefont {Z.-b.}\ \bibnamefont {Li}}, \bibinfo {author} {\bibfnamefont {Z.}~\bibnamefont {Zhang}}, \bibinfo {author} {\bibfnamefont {Y.-z.}\ \bibnamefont {Fu}}, \bibinfo {author} {\bibfnamefont {C.}~\bibnamefont {Wang}}, \bibinfo {author} {\bibfnamefont {L.}~\bibnamefont {Wei}}, \ and\ \bibinfo {author} {\bibfnamefont {S.}~\bibnamefont {Bai}},\ }\bibfield  {title} {\enquote {\bibinfo {title} {High performance and exchange coupling in magnetization reversal of sintered (nd, dy)-fe-b magnets},}\ }\href@noop {} {\bibfield  {journal} {\bibinfo  {journal} {J. Alloys Compd.}\ }\textbf {\bibinfo {volume} {926}},\ \bibinfo {pages} {166944} (\bibinfo {year} {2022})}\BibitemShut {NoStop}%
\bibitem [{\citenamefont {Aharoni}(1998)}]{RN1406}%
  \BibitemOpen
  \bibfield  {author} {\bibinfo {author} {\bibfnamefont {A.}~\bibnamefont {Aharoni}},\ }\bibfield  {title} {\enquote {\bibinfo {title} {Demagnetizing factors for rectangular ferromagnetic prisms},}\ }\href@noop {} {\bibfield  {journal} {\bibinfo  {journal} {J. Appl. Phys.}\ }\textbf {\bibinfo {volume} {83}},\ \bibinfo {pages} {3432} (\bibinfo {year} {1998})}\BibitemShut {NoStop}%
\bibitem [{\citenamefont {Becker}(1971)}]{RN1525}%
  \BibitemOpen
  \bibfield  {author} {\bibinfo {author} {\bibfnamefont {J.~J.}\ \bibnamefont {Becker}},\ }\bibfield  {title} {\enquote {\bibinfo {title} {Magnetization discontinuities in cobalt-rare-earth particles},}\ }\href@noop {} {\bibfield  {journal} {\bibinfo  {journal} {J. Appl. Phys.}\ }\textbf {\bibinfo {volume} {42}},\ \bibinfo {pages} {1537} (\bibinfo {year} {1971})}\BibitemShut {NoStop}%
\bibitem [{\citenamefont {Xing}\ \emph {et~al.}(2020)\citenamefont {Xing}, \citenamefont {Shen}, \citenamefont {Chen}, \citenamefont {Huang}, \citenamefont {Gao}, \citenamefont {Zheng}, \citenamefont {Zhang}, \citenamefont {Li}, \citenamefont {Hu}, \citenamefont {Qian}, \citenamefont {Cao}, \citenamefont {Zhang}, \citenamefont {Fan}, \citenamefont {Ma}, \citenamefont {Wang}, \citenamefont {Yin}, \citenamefont {Lei}, \citenamefont {Ji}, \citenamefont {Du}, \citenamefont {Yang}, \citenamefont {Wang}, \citenamefont {Shen}, \citenamefont {Lin}, \citenamefont {Liu}, \citenamefont {Shen}, \citenamefont {Wang},\ and\ \citenamefont {Gao}}]{RN1230}%
  \BibitemOpen
  \bibfield  {author} {\bibinfo {author} {\bibfnamefont {Y.}~\bibnamefont {Xing}}, \bibinfo {author} {\bibfnamefont {J.}~\bibnamefont {Shen}}, \bibinfo {author} {\bibfnamefont {H.}~\bibnamefont {Chen}}, \bibinfo {author} {\bibfnamefont {L.}~\bibnamefont {Huang}}, \bibinfo {author} {\bibfnamefont {Y.}~\bibnamefont {Gao}}, \bibinfo {author} {\bibfnamefont {Q.}~\bibnamefont {Zheng}}, \bibinfo {author} {\bibfnamefont {Y.~Y.}\ \bibnamefont {Zhang}}, \bibinfo {author} {\bibfnamefont {G.}~\bibnamefont {Li}}, \bibinfo {author} {\bibfnamefont {B.}~\bibnamefont {Hu}}, \bibinfo {author} {\bibfnamefont {G.}~\bibnamefont {Qian}}, \bibinfo {author} {\bibfnamefont {L.}~\bibnamefont {Cao}}, \bibinfo {author} {\bibfnamefont {X.}~\bibnamefont {Zhang}}, \bibinfo {author} {\bibfnamefont {P.}~\bibnamefont {Fan}}, \bibinfo {author} {\bibfnamefont {R.}~\bibnamefont {Ma}}, \bibinfo {author} {\bibfnamefont {Q.}~\bibnamefont {Wang}}, \bibinfo {author} {\bibfnamefont {Q.}~\bibnamefont {Yin}}, \bibinfo {author} {\bibfnamefont {H.}~\bibnamefont {Lei}}, \bibinfo {author} {\bibfnamefont {W.}~\bibnamefont {Ji}}, \bibinfo {author} {\bibfnamefont {S.}~\bibnamefont {Du}}, \bibinfo {author} {\bibfnamefont {H.}~\bibnamefont {Yang}}, \bibinfo {author} {\bibfnamefont {W.}~\bibnamefont {Wang}}, \bibinfo {author} {\bibfnamefont {C.}~\bibnamefont {Shen}}, \bibinfo {author} {\bibfnamefont {X.}~\bibnamefont {Lin}}, \bibinfo {author} {\bibfnamefont {E.}~\bibnamefont {Liu}}, \bibinfo {author} {\bibfnamefont {B.}~\bibnamefont {Shen}}, \bibinfo {author} {\bibfnamefont {Z.}~\bibnamefont {Wang}}, \ and\ \bibinfo {author} {\bibfnamefont {H.~J.}\ \bibnamefont {Gao}},\ }\bibfield  {title} {\enquote {\bibinfo {title} {Localized spin-orbit polaron in magnetic weyl semimetal co$_3$sn$_2$s$_2$},}\ }\href@noop {} {\bibfield  {journal} {\bibinfo  {journal} {Nat. Commun.}\ }\textbf {\bibinfo {volume} {11}},\ \bibinfo {pages} {5613} (\bibinfo {year} {2020})}\BibitemShut {NoStop}%
\end{thebibliography}%

\end{document}